\documentclass[onecolumn,secnumarabic,amssymb, nofootinbib,notitlepage,aps, pra]{revtex4-1}

\usepackage{amsthm}
\usepackage{amsmath}
\usepackage{color}
\usepackage{graphicx}
\usepackage{hyperref}
\usepackage{array}
\usepackage{subfig}

\newcommand{\ket}[1]{\left| #1\right. \rangle}
\newcommand{\bra}[1]{\langle\left. #1 \right|}
\newcommand{\braket}[2]{\langle\left. #1 \right| #2 \rangle }
\newcommand{\corr}[1]{\langle #1 \rangle}
\newcommand{\id}{\mathbb{I}}
\newcommand{\dd}{\mathrm{d}}

\newcommand{\figref}[1]{Fig. \ref{#1}}
\newcommand{\secref}[1]{Section \ref{#1}}
\newcommand{\eqnref}[1]{(\ref{#1})}

\newcommand{\includepicture}[1]{
\begin{tabular}{c} 
\includegraphics[width=0.15\columnwidth]{#1}
\end{tabular}}

\begin{document}

\title{Experimental weak measurement of two non-commuting observables}%

\author{Gilles P\"utz}%
\email[]{these authors contributed equally to this work}
\affiliation{Group of Applied Physics, University of Geneva, CH-1211 Geneva 4, Switzerland.}
\affiliation{Institute for Theoretical Physics, ETH-Z\"urich, CH-8093, Z\"urich, Switzerland.}
\author{Tomer Barnea}%
\email[]{these authors contributed equally to this work}
\affiliation{Group of Applied Physics, University of Geneva, CH-1211 Geneva 4, Switzerland.}
\author{Nicolas Gisin}%
\affiliation{Group of Applied Physics, University of Geneva, CH-1211 Geneva 4, Switzerland.}
\author{Anthony Martin}%
\email[]{anthony.martin@unige.ch}
\affiliation{Group of Applied Physics, University of Geneva, CH-1211 Geneva 4, Switzerland.}

\date{\today}%
\begin{abstract}
The fact that not all quantum observables are jointly measurable is one of the major differences between quantum and classical theory. In the former, non-commuting observables can only be simultaneously measured with limited precision. We report on an experimental implementation of such a simultaneous measurement of two non-commuting observables based on the framework of weak von Neumann measurements. We use a photonic setup where the polarisation degree of freedom acts as the system and the two components of the transversal position correspond to the pointers of the measurement apparatuses. In addition, the theory shows that these weak measurements demonstrate a counter-intuitive non-monotonicity: weaker measurements can potentially reveal more information than stronger ones.
\end{abstract}

\maketitle

\section{Introduction}
\label{sec:intro}

One of the big differences between quantum and classical theory is that in the former, some observables cannot be measured together. This is due to the fact that measurements perturb the state of the system that is being measured. The conventional wisdom is that two observables cannot be measured together unless they commute. The more precise statement would be that one cannot perfectly measure two non-commuting observables together. It is, however, possible to do it by accepting some limited precision on each observable.\\

One of the most famous examples is given by the impossibility to simultaneously measure position and momentum~\cite{heisenberg27}. Arthurs and Kelly~\cite{arthurs65} proposed a way of implementing a weak simultaneous measurement of these two observables based on the von Neumann~\cite{neumann55} measurement model. The von Neumann model performs measurements by coupling the observable of a system one wishes to measure to a pointer, representing the measurement apparatus, that is then itself measured projectively by simply reading off the value it points to. If the pointer is very broad compared to the distance between the eigenvalues of the observable or if the coupling between the system and the pointer is very weak, then some measurement precision is lost. Note however that this is not the same as simply performing a noisy measurement; in fact by performing this measurement with a broad pointer or a weak coupling, we achieve that the post-measurement state of the system is less perturbed than it would be by a projective measurement. It thus constitutes a weak measurement.

Arthurs and Kelly extended this scheme to using two pointers and coupling two different observables, in their case position and momentum, to each. Their model was further developed and investigated in various theoretical works, e.g.~\cite{appleby98,lorenzo13,bullock14}, and then refined to measuring the direction of a spin~\cite{dariano02}, i.e. to measuring three non-commuting spin-observables simultaneously by coupling them to three independent pointers.\\

In the present work, we report on a proof of principle experiment implementing a simultaneous measurement of the $\sigma_z$ and $\sigma_x$ observables of a qubit encoded in photon polarisation. Let us stress that our experiment does not rely on post-selection or weak value amplification, as opposed to the scheme of Mitchison et al.~\cite{mitchison07}, which was experimentally implemented by Piacentini et al.~\cite{piacentini15} and relies on weak-value amplification~\cite{aharonov88,pryde05} through post-selection (see e.g.~\cite{resch04,hosten08}). We use birefringent crystals to couple the polarization degree of freedom of photons, which we prepare in different initial states, to their transversal spatial mode, making them the system and the pointers, respectively. After measuring the position using a CCD camera, we provide guesses for the desired observables of the system based on the measurement result and compare them to the actual value of these observables of the input polarization state. Similar attempts to conduct state tomography have been conducted in different systems, e.g.~\cite{shay16} where they use superconducting qubits and focus on the dynamics of the process or~\cite{howland14} with compressive sensing techniques. The details of our implementation and the analysis of the data is presented in \secref{sec:experiment}.\\

Our results display a surprising phenomenon: as opposed to weak measurements of a single observable, where a weaker measurement leads to less information gain~\cite{poulin05}, we observe a non-monotonicity when weakly measuring two (or more~\cite{barnea16}) observables. In some regimes, making the measurement weaker by weakening the coupling or broadening the pointer reveals more information. This effect is not yet fully understood and we discuss it from a theoretical point of view in \secref{sec:theory}, where we also present the theoretical background of the experiment.

\section{Theoretical background}
\label{sec:theory}
We describe a measurement apparatus in the von Neumann model\cite{neumann55} by a (continuous) pointer which is initially in some state 
\begin{align*}
\ket{\Phi_{i}}=\int_{-\infty}^{+\infty}\dd xG_{\Delta}(x)\ket{x},
\end{align*}
where $\ket{x}$ is the position basis and $G_{\Delta}(x)=\frac{1}{\sqrt{2\pi}\Delta}e^{-\frac{x^2}{4\Delta^2}}$ is a Gaussian distribution of spread $\Delta$. A measurement is performed by coupling the observable $\hat{O}$ of a system to this pointer and then projectively measuring the position of the pointer. The coupling is achieved by the interaction Hamiltonian
\begin{align*}
H_{int}(t)=g(t)\hat{O}\otimes\hat{p},
\end{align*}
where $\hat{p}$ is the displacement operator of the pointer and $g(t)$ describes the temporal implementation of the interaction. Denoting by $\ket{\psi_i}$ the initial state of the system, we have that the state of system and pointer after the interaction is given by
\begin{align*}
\ket{\Psi_f}&=e^{-i\int_{-\infty}^{\infty}\dd t H_{int}(t)}\left(\ket{\psi_i}\otimes\ket{\Phi_i}\right)\\
&=e^{-i\delta\hat{O}\otimes\hat{p}}\left(\ket{\psi_i}\otimes\ket{\Phi_i}\right)
\end{align*}
where we defined $\delta=\int_{-\infty}^{\infty}\dd tg(t)$. After the interaction, we measure the position of the pointer and from it deduce information about the observable $\hat{O}$. The strength of the measurement depends on the strength of the interaction $\delta$ and the spread of the pointer $\Delta$. Intuitively, a stronger coupling and a narrower pointer lead to more information~\cite{poulin05}.

Within this framework it is quite intuitive that in order to measure two (or more) different observables simultaneously, one can simply employ additional pointers and couplings. For non-commuting observables, the information that can be revealed this way is of course limited. We employ this technique for the specific case of measuring the $\sigma_z$ and $\sigma_x$ observables\footnote{$\sigma_z=\ket{0}\bra{0}-\ket{1}\bra{1}$ and $\sigma_x=\left(\frac{\ket{0}+\ket{1}}{\sqrt{2}}\frac{\bra{0}+\bra{1}}{\sqrt{2}}\right)-\left(\frac{\ket{0}-\ket{1}}{\sqrt{2}}\frac{\bra{0}-\bra{1}}{\sqrt{2}}\right)$ are the Pauli matrices.} of a qubit in the $x-z$ plane using Gaussian pointers. In other words, the initial state of the system is given by $\ket{\psi_i}=\cos\frac{\theta_i}{2}\ket{0}+\sin\frac{\theta_i}{2}\ket{1}$, the initial state of the pointers by
\begin{align}
\label{eq:initialpointer}
\ket{\Phi_i}=\int_{-\infty}^{+\infty}\dd zG_{\Delta}(z)\ket{z}\otimes\int_{-\infty}^{+\infty}\dd x G_{\Delta}(x)\ket{x}
\end{align}
where for simplicity we only consider pointers of equal spread $\Delta$. The interaction Hamiltonian is given by
\begin{align}
\label{eq:Hint}
H_{int}(t)=g(t)\left(\sigma_z\otimes\hat{p}_z\otimes\id+\sigma_x\otimes\id\otimes\hat{p}_x\right).
\end{align}
Similar to the single pointer scenario, the post-interaction state is given by
\begin{align}
\ket{\Psi_f}=e^{-i\delta \left(\sigma_z\otimes\hat{p}_z\otimes\id+\sigma_x\otimes\id\otimes\hat{p}_x\right)}\left(\ket{\psi_i}\otimes\ket{\Phi_i}\right).
\label{eq:psifinal}
\end{align}
We projectively measure the position of the pointers. The probability of finding the first pointer at $z$ and the second at $x$ given that the initial state was $\ket{\psi_i}$ evaluates to
\begin{align*}
P(z,x|\psi_i)=\text{tr}\left((\id\otimes\ket{z}\bra{z}\otimes\ket{x}\bra{x}) \ket{\Psi_f}\bra{\Psi_f}\right).
\end{align*}
We present an explicit computation of $P(z,x|\psi_i)$ in Appendix \ref{app:meas}.
We plot $P(z,x|\psi_i)$ in the first column of \figref{fig:images} for $\theta_i=0$ and $\theta_i=-\frac{\pi}{2}$ and different $\Delta/\delta$. It can be seen that given the full picture, the states are clearly distinguishable. The fact that the probability distributions overlap eliminates this distinguishability in the single-shot case. \\



Our goal is to determine the expectation values of the observables $\sigma_z$ and $\sigma_x$ of the initial qubit state $\ket{\psi_i}$, given by $\corr{\sigma_{z/x}}_{\ket{\psi_i}}=\bra{\psi_i}\sigma_{z/x}\ket{\psi_i}$. We thus have to deduce their values from our measurement, i.e. the positions of the two pointers, denoted by $z=r\cos\theta_g$ and $x=r\sin\theta_g$. Since the measurement is imprecise, we can only give guesses for these expectation values, which we denote  $\corr{\sigma_{z/x}}_g$. In this work, we use the more or less straightforward guess given by $\corr{\sigma_z}_g=\cos\theta_g$ and $\corr{\sigma_x}_g=\sin\theta_g$, i.e. we interpret the two pointers as one two-dimensional pointer and deduce $\corr{\sigma_{z/x}}_g$ from its direction. 

We can write any 1-qubit density matrix in the $x-z$ plane as follows
\begin{align}
\rho=\frac{1}{2}\left(\id+\corr{\sigma_z}_{\rho}\sigma_z+\corr{\sigma_x}_{\rho}\sigma_x\right).
\label{eq:rho}
\end{align}
In the case of a pure state, i.e. $\rho=\ket{\psi}\bra{\psi}$, we have that $\corr{\sigma_z}_{\ket{\psi}}^2+\corr{\sigma_x}_{\ket{\psi}}^2=1$. The expectation values of $\sigma_z$ and $\sigma_x$ thus contain all the relevant information; if we were able to measure both $\corr{\sigma_z}_{\ket{\psi_i}}$ and $\corr{\sigma_x}_{\ket{\psi_i}}$ perfectly, then we could fully reconstruct our initial state. A relevant figure of merit is therefore given by the fidelity between the initial state $\ket{\psi_i}$ and the state $\rho_g$ we guess based on our measurement (which can easily be derived from \eqref{eq:rho} by simply changing $\corr{\sigma_{z/x}}_{\rho}$ to $\corr{\sigma_{z/x}}_{g}$). For the guess we examine in this work, the state $\rho_g$ is pure and can be written as $\ket{\psi_g(z,x)}=\cos\frac{\theta_g}{2}\ket{0}+\sin\frac{\theta_g}{2}\ket{1}$. The average fidelity is then given by
\begin{align*}
F_{avg}(\ket{\psi_i})&=\int_{-\infty}^{+\infty}\dd z\int_{-\infty}^{+\infty}\dd xP(z,x|\psi_i)|\braket{\psi_{g}(z,x)}{\psi_i}|^2\\
&=\int_0^{2\pi}\dd\theta_g\left(\int_{0}^{+\infty}\dd rrP(r\cos\theta_g,r\sin\theta_g|\psi_i)\right)\cos^2\left(\frac{\theta_i-\theta_g}{2}\right)
\end{align*}

\begin{figure}
\includegraphics[width=0.4\textwidth]{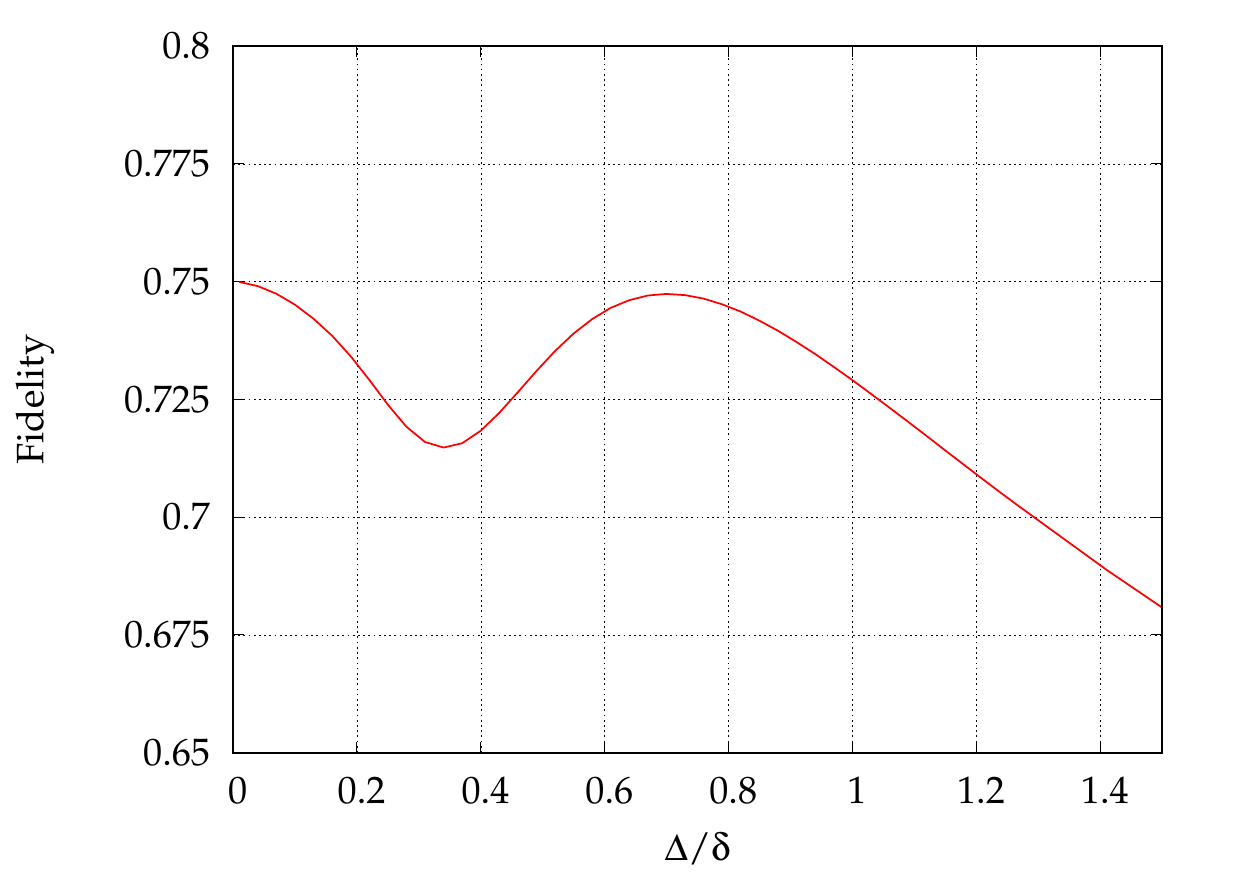}
\caption{Fidelity of the guessed state as a function of the strength of the measurement $\Delta/\delta$. In the beginning, the fidelity behaves as expected, i.e. it goes down as the measurement gets weaker. However, there is a second-wind-effect since the fidelity climbs back up to almost the threshold value at $\Delta/\delta\approx 0.7$.}
\label{fig:fullFidelity}
\end{figure}

As shown in \figref{fig:fullFidelity}, when $\Delta/\delta$ is close to 0, the average fidelity approaches $\frac{3}{4}$, which is the maximal possible value in quantum mechanics~\cite{buscemi05}. Our measurement and guess are thus optimal in the strong coupling regime. As the measurement gets weaker, it could be expected that the fidelity decreases steadily. This is however not the case; instead we observe a "second wind" effect where after an initial decline the fidelity goes back up to close to its maximal value before before dropping off to approach $\frac{1}{2}$ asymptotically.

Developing an intuition for this rather strange effect is left as an open question. We conjecture that it results from the tradeoff between information gain and perturbance: while weaker measurements reveal less information, they also perturb less and thereby let the other measurement extract potentially more information. The exact nature of this tradeoff for the case of simultaneous measurements is yet to be understood.

In the following, we report on a first approximate photonic implementation of the measurement presented here.

\section{Experimental implementation}
\label{sec:experiment}

From an experimental point of view it is not obvious how to implement the Hamiltonian given in Eq.~\eqref{eq:Hint}. However the time evolution to yield the final state seen in Eq.~\eqref{eq:psifinal} can be reformulated using the well-known Trotter formula~\cite{trotter59}
\begin{align}
\label{eq:Trotter}
e^{-i\delta\left(\sigma_z\otimes\hat{p}_z\otimes\id+\sigma_x\otimes\id\otimes\hat{p}_x\right)}=\lim_{n\rightarrow\infty}\left(e^{-i\frac{\delta}{n}\sigma_x\otimes\id\otimes\hat{p}_x}e^{-i\frac{\delta}{n}\sigma_z\otimes\hat{p}_z\otimes\id}\right)^n.
\end{align}
Alternately coupling to $\sigma_z$ and $\sigma_x$ in succession can thus implement the desired evolution. 

In the special case of a qubit encoded in the polarization degree of freedom of a photon, these couplings can be implemented by using birefringent beam displacers (BBD).
The BBDs introduce a polarization dependent spatial displacement in the plane orthogonal to the propagation axis by moving photons of one polarization by a distance $d$ and letting those of the orthogonal polarization pass straight through. In other words, if we turn a BBD crystal by an angle $\theta$ around the propagation axis, then the transformation that they implement is given by $e^{-i\frac{d}{2}\sigma_{\theta}\otimes\hat{p}_{\theta}}$ where $\sigma_{\theta}=\ket{\theta}\bra{\theta}-\ket{\theta+\pi}\bra{\theta+\pi}$ with $\ket{\theta}=\cos\frac{\theta}{2}\ket{0}+\sin\frac{\theta}{2}\ket{1}$ and $\hat{p}_{\theta}$ is the spatial displacement operator in the direction given by the polar angle $\theta$.\footnote{Note that we use here that displacing one polarization by $d$ and leaving the other is the same as displacing one by $\frac{d}{2}$ in one direction and the other by the same distance in the opposite direction. This simply corresponds to shifting the reference frame.}

\begin{figure}
\includegraphics[width=0.4\textwidth]{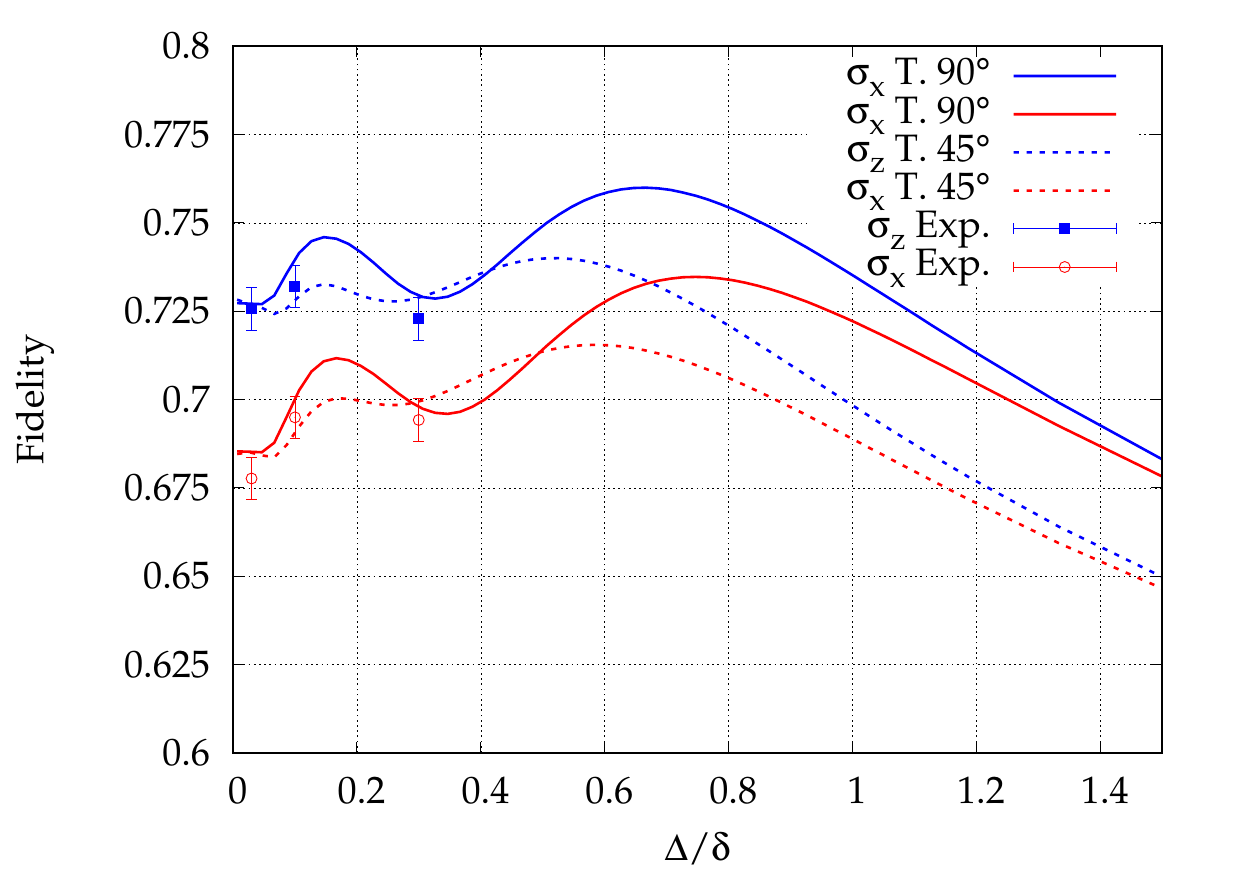}
\caption{Fidelity of the guessed state as a function of the strength of the measurement $\Delta/\delta$ for the Trotter measurement as well as for the implementation with birefringent crystals for $n=6$. While the fidelities are the same for $\ket{H}$ and $\ket{V}$ as well as for $\ket{D}=\frac{1}{\sqrt{2}}\left(\ket{H}+\ket{V}\right)$ and $\ket{A}=\frac{1}{\sqrt{2}}\left(\ket{H}-\ket{V}\right)$, they differ between the two bases. Strong measurements are no longer optimal for this type of measurements. The points correspond to the fidelity obtained from the experimental data given in the fourth column of \figref{fig:images}.}
\label{fig:simFidelity}
\end{figure}

\begin{figure}
\begin{tabular}{c c}
$\frac{\Delta}{\delta}$ = 0.03&
\begin{tabular}{c c c c c}
&Continuous & Trotter 90$^\circ$ & Trotter 45$^\circ$\ & Measured\\
$\ket{0}$ & \includepicture{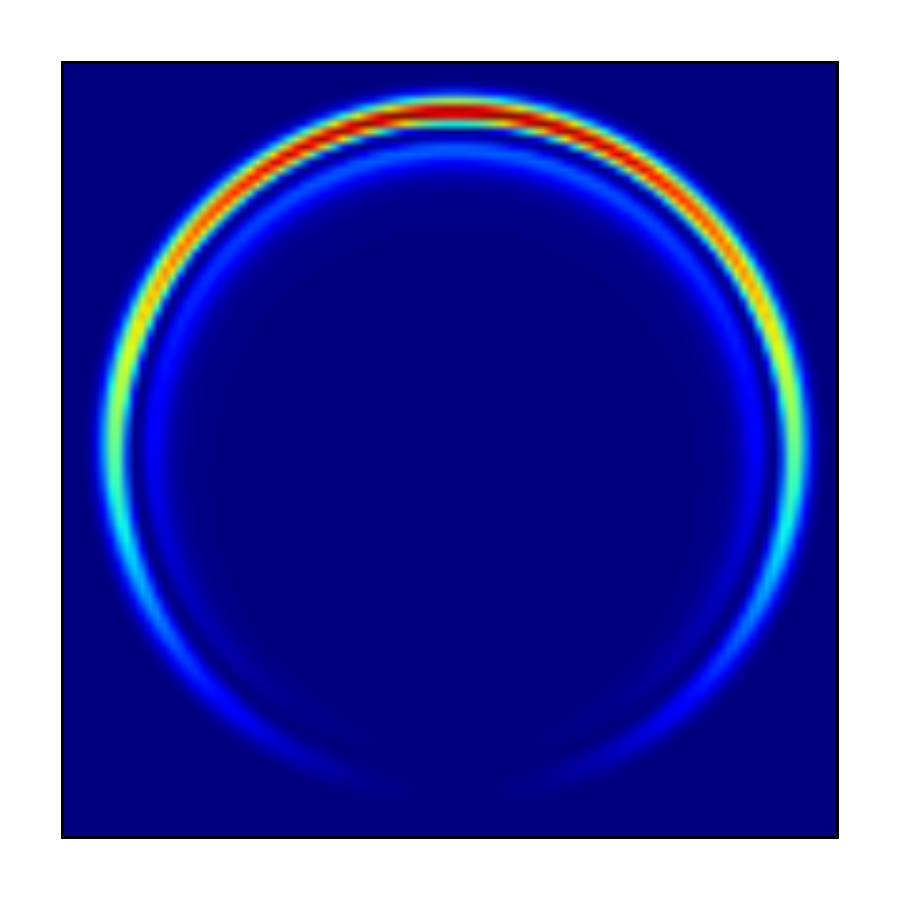} & \includepicture{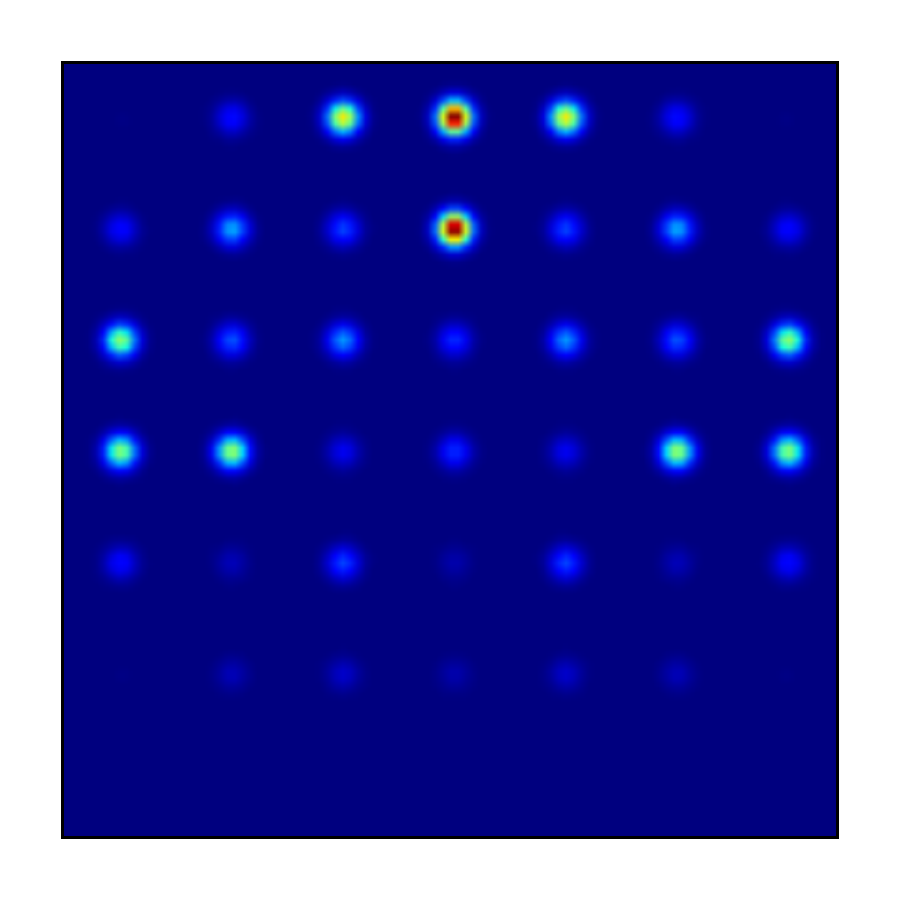} & \includepicture{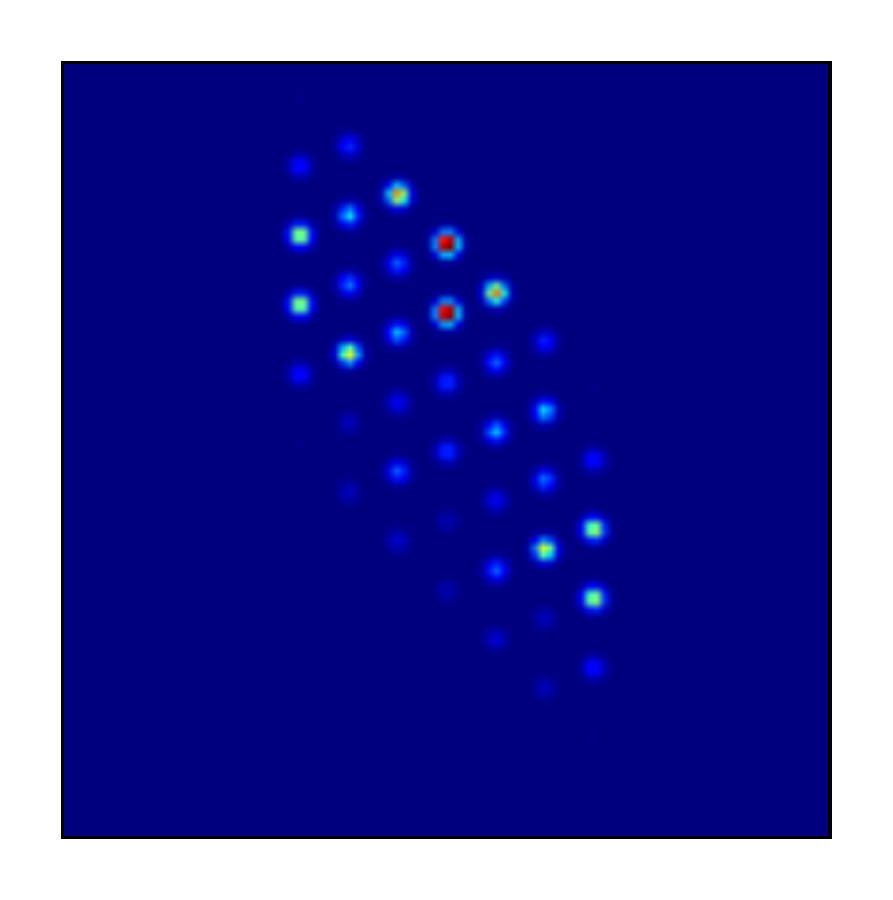} & \includepicture{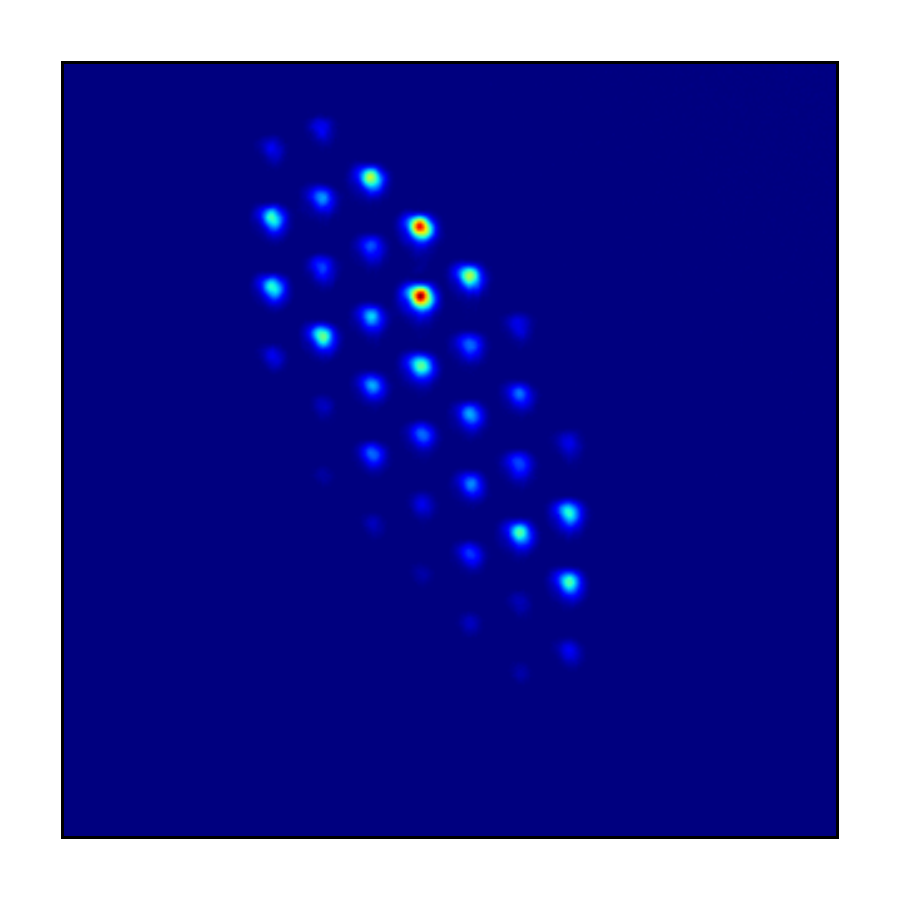}\\
$\frac{\ket{0}-\ket{1}}{\sqrt{2}}$ & \includepicture{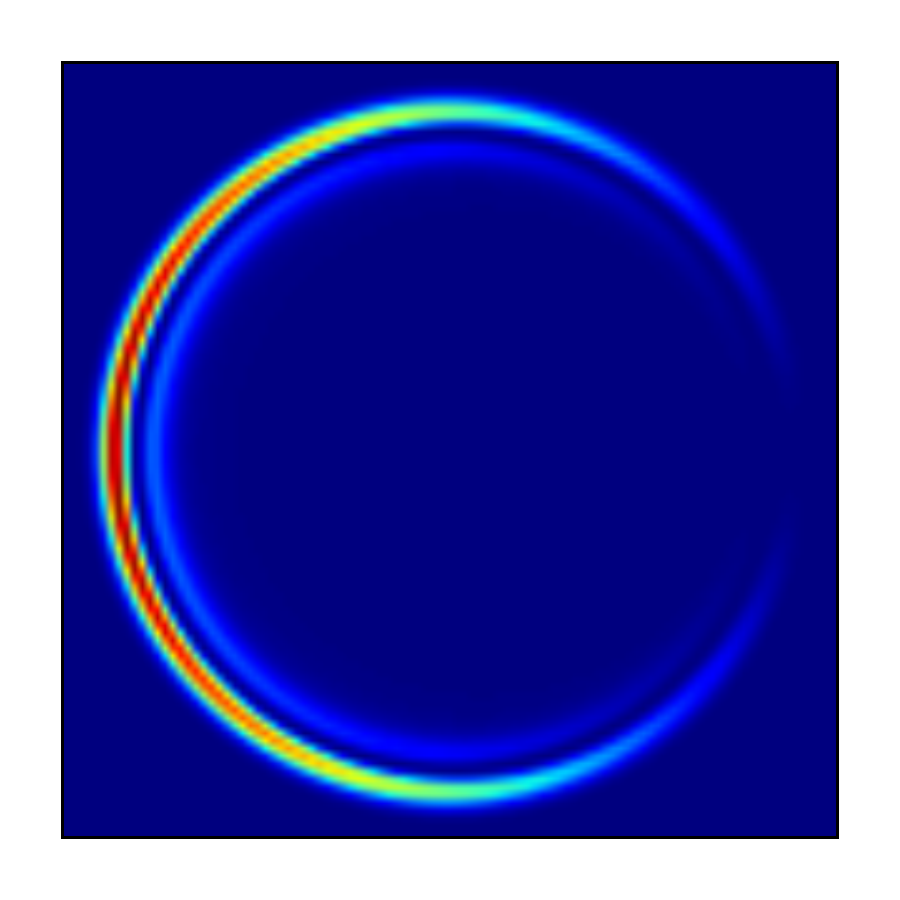} & \includepicture{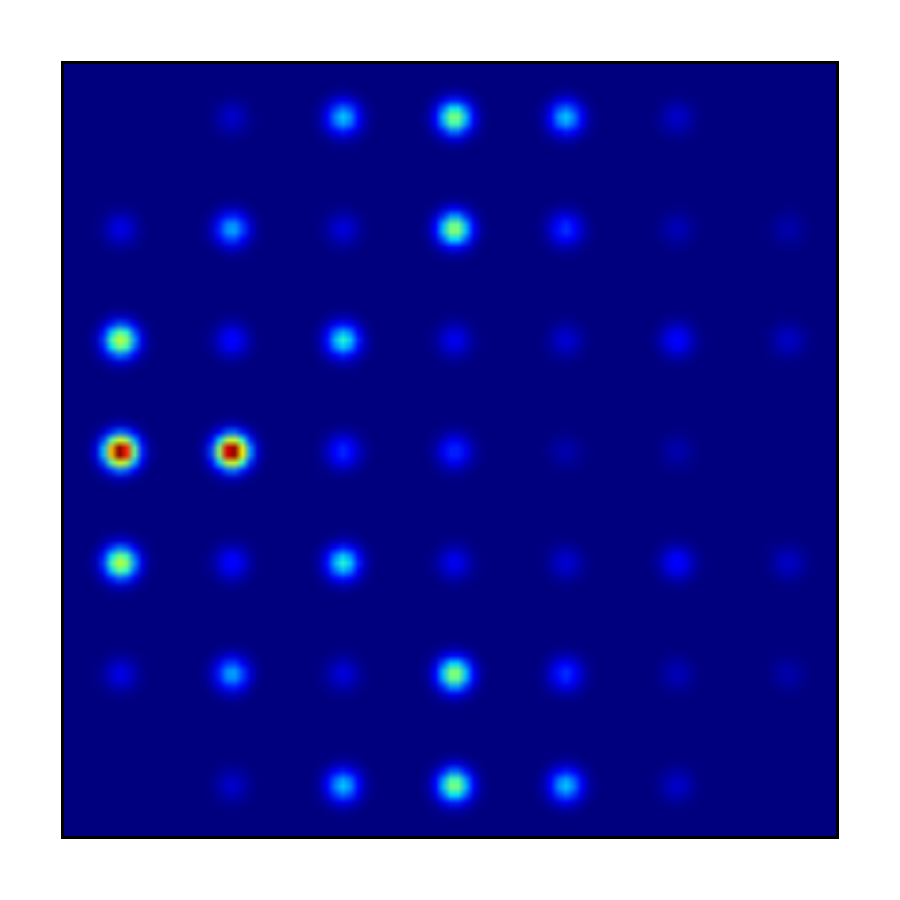} & \includepicture{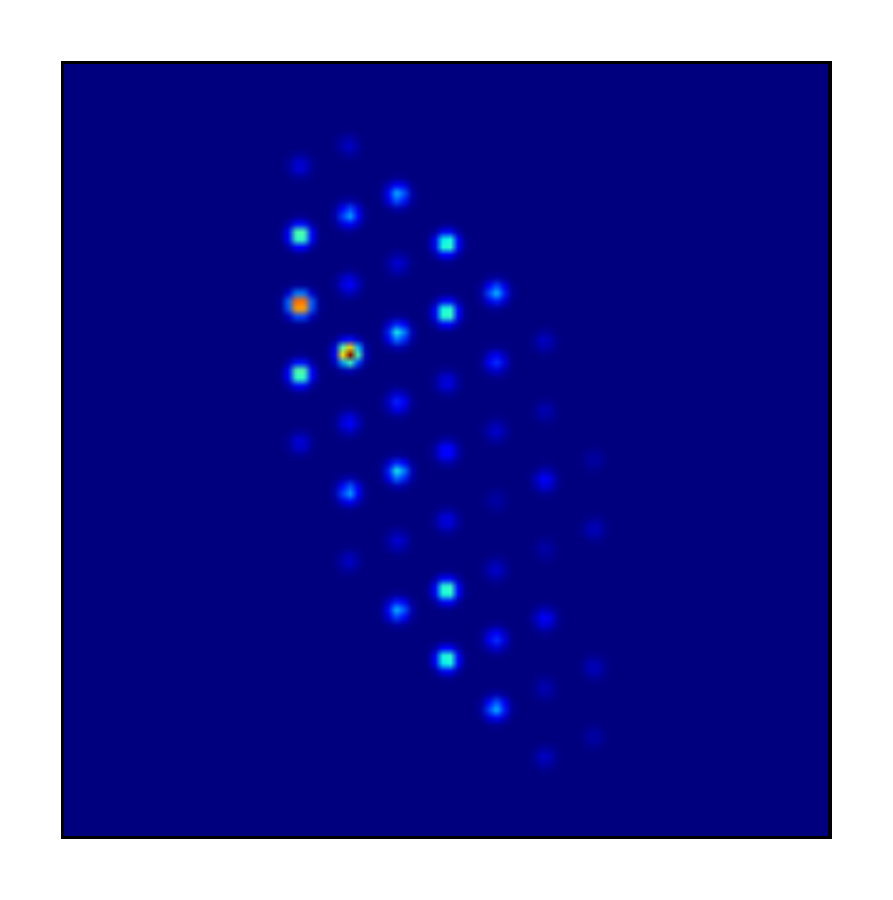} & \includepicture{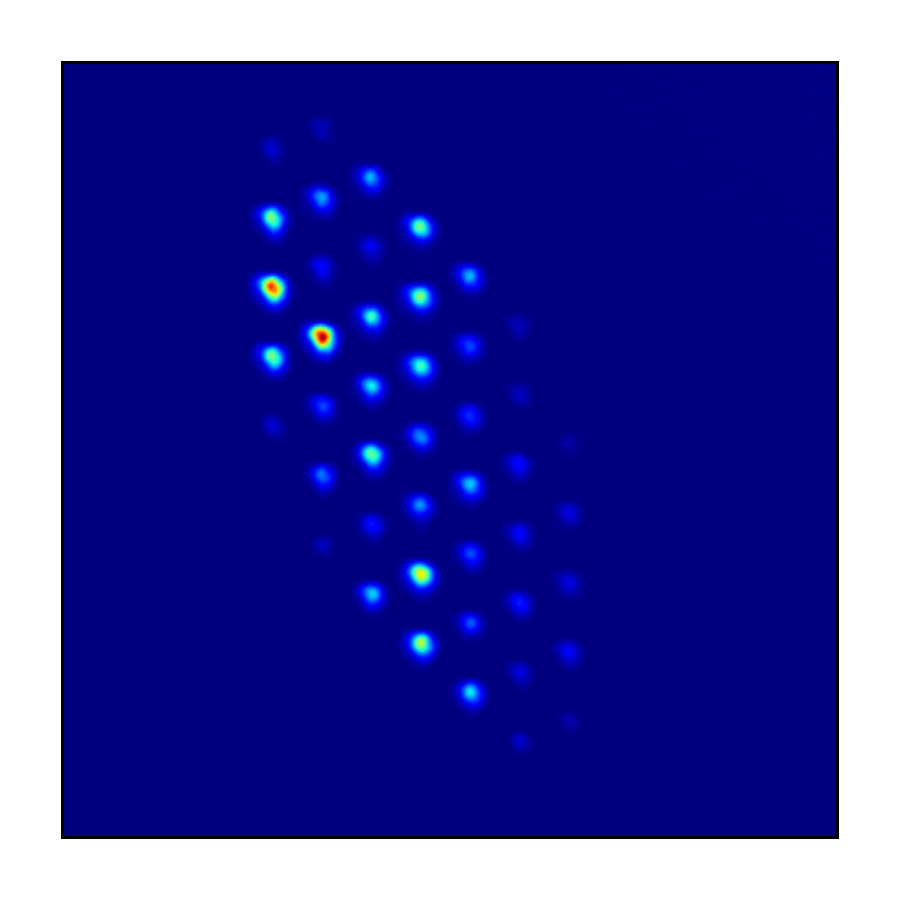}\\
\end{tabular} \\
\hline \\
$\frac{\Delta}{\delta}$ = 0.15&
\begin{tabular}{c c c c c}
$\ket{0}$  & \includepicture{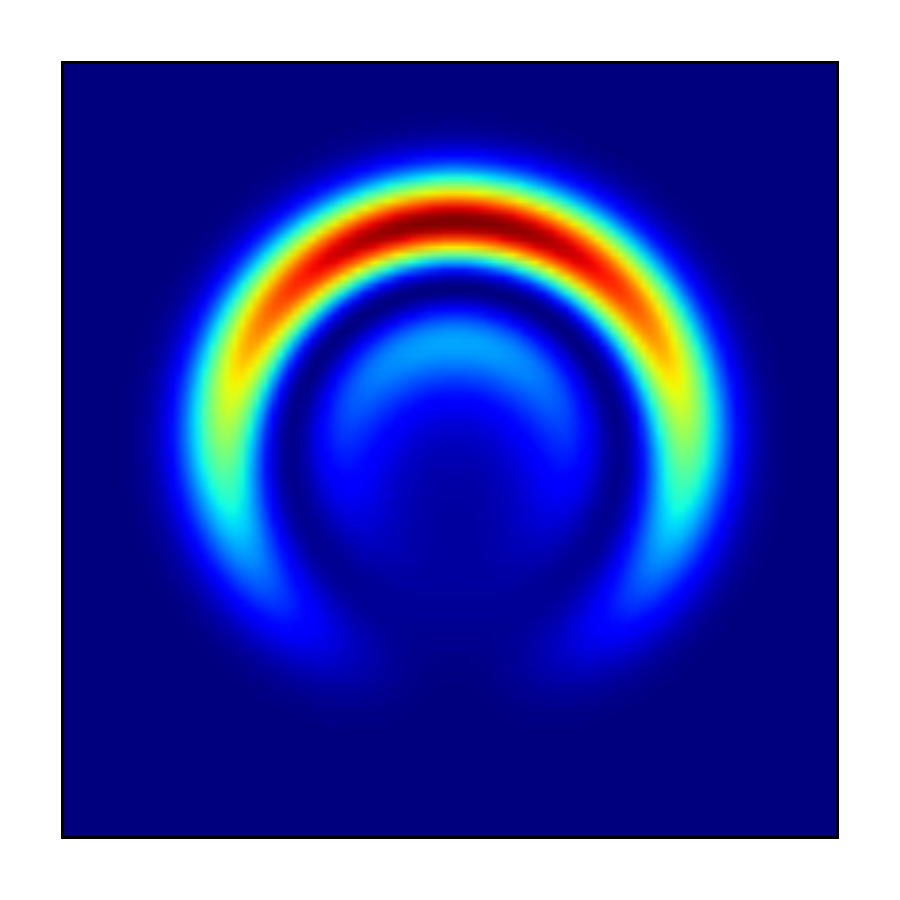} & \includepicture{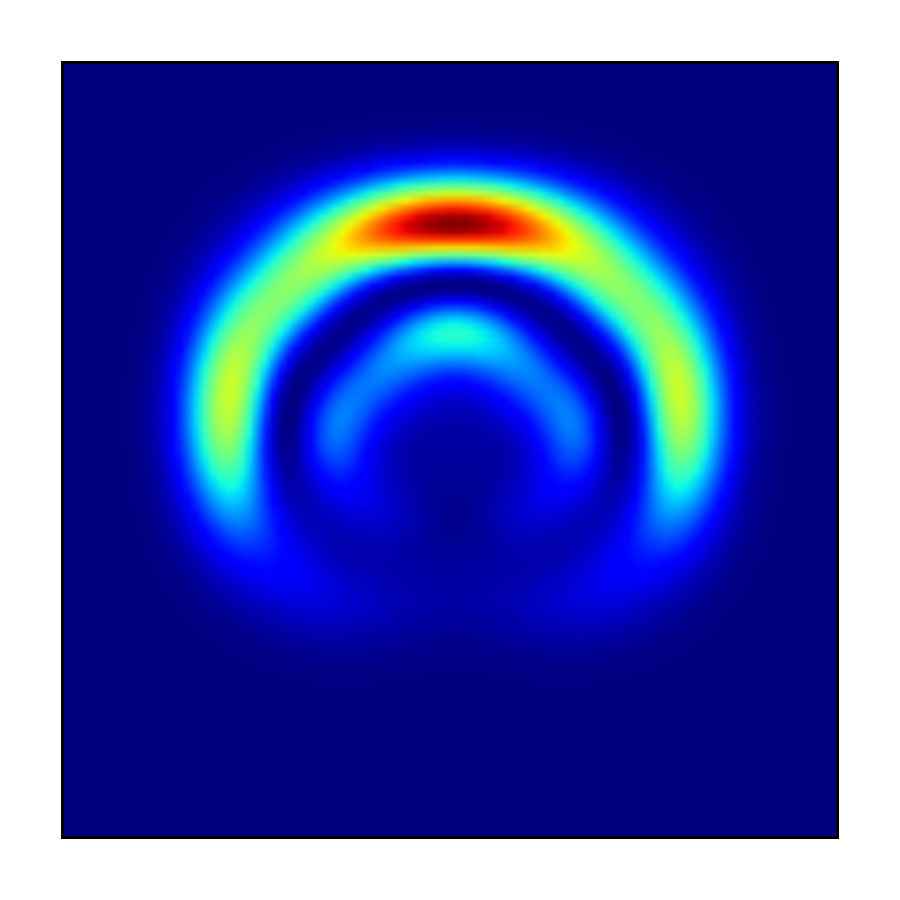} & \includepicture{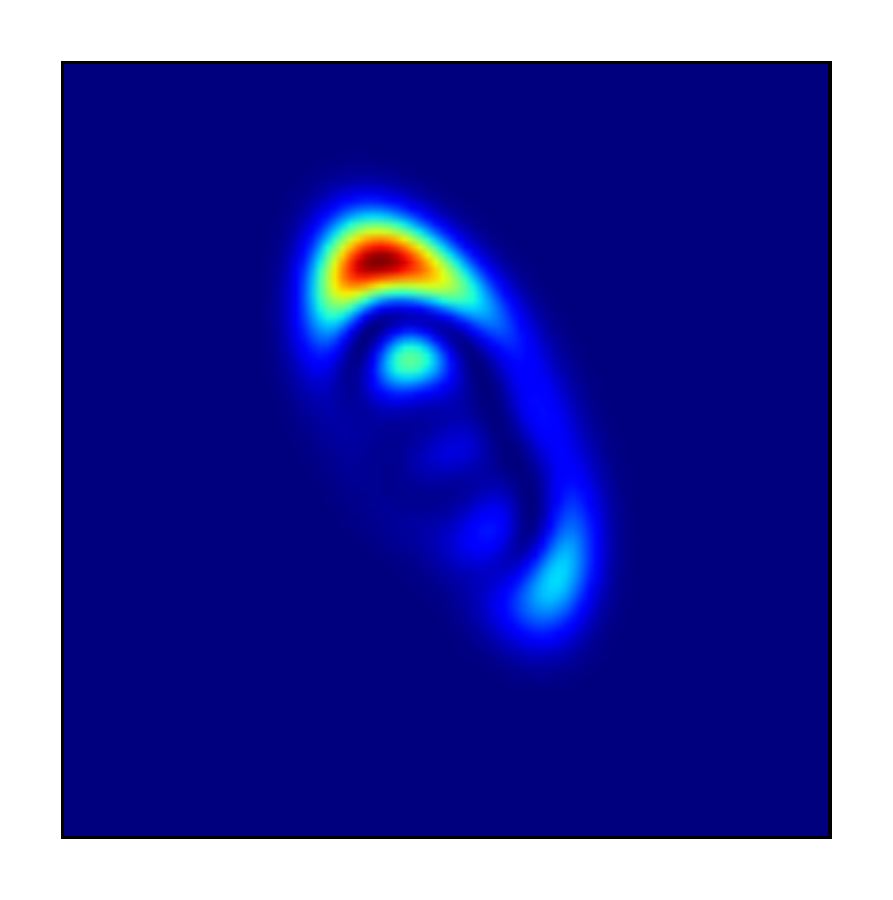} & \includepicture{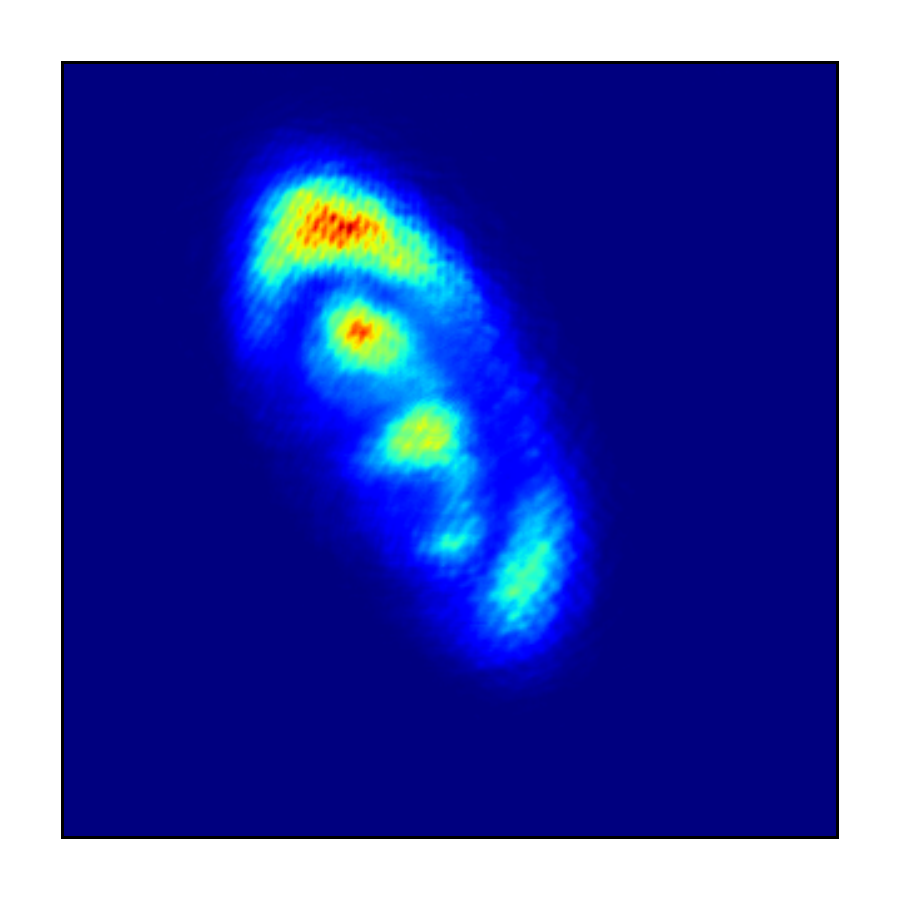}\\
$\frac{\ket{0}-\ket{1}}{\sqrt{2}}$ & \includepicture{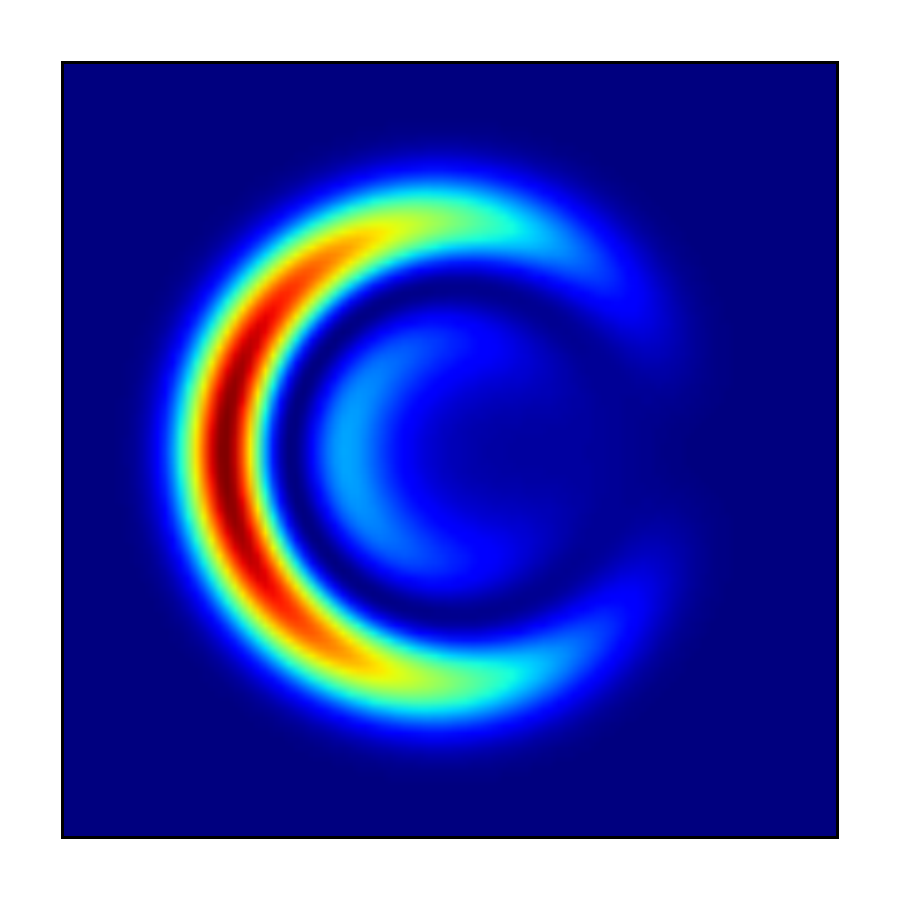} & \includepicture{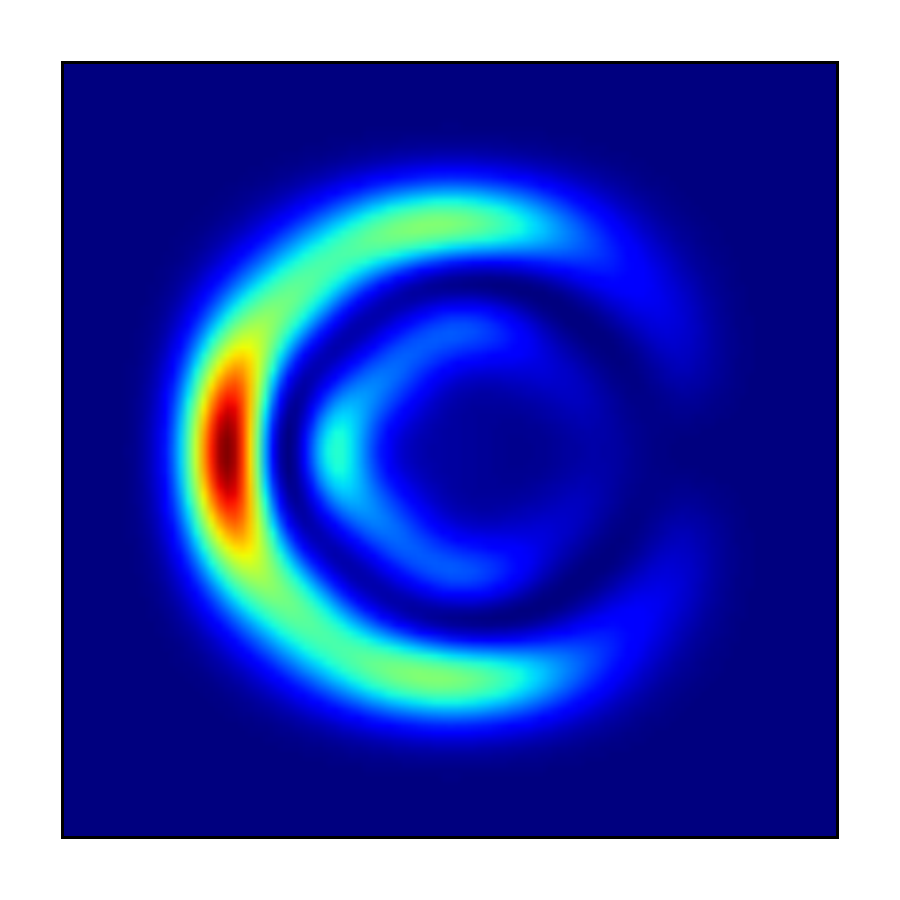} & \includepicture{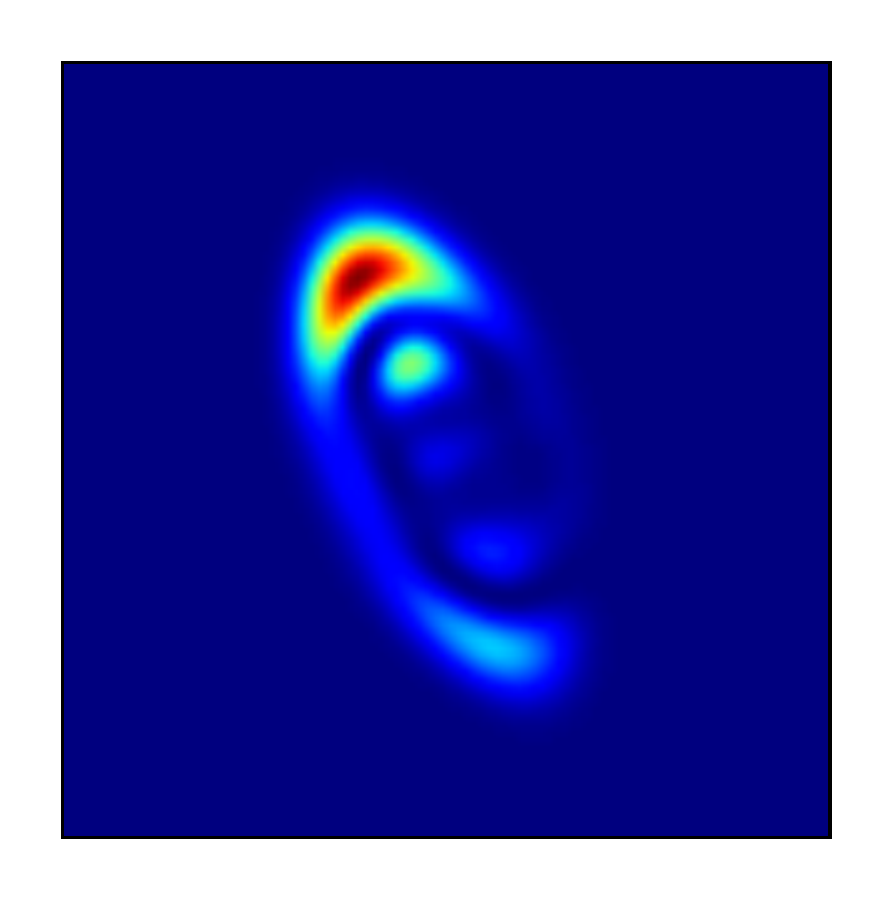} & \includepicture{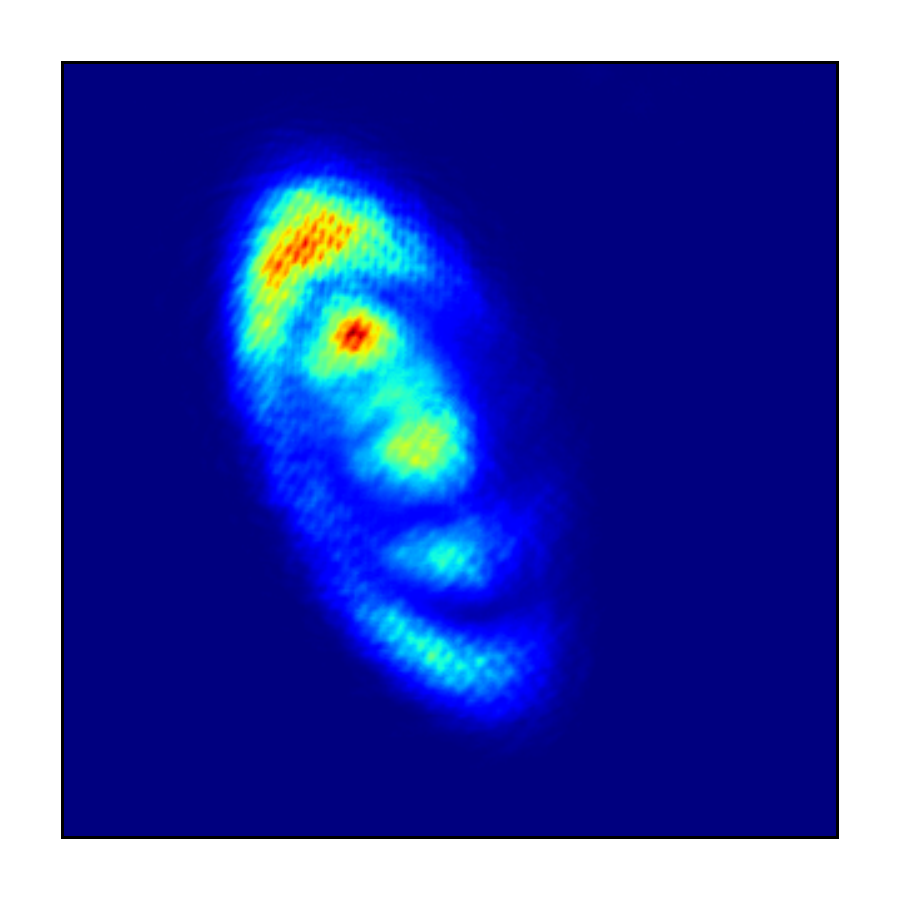}\\
\end{tabular} \\
\hline \\
$\frac{\Delta}{\delta}$ = 0.3&
\begin{tabular}{c c c c c}
$\ket{0}$  & \includepicture{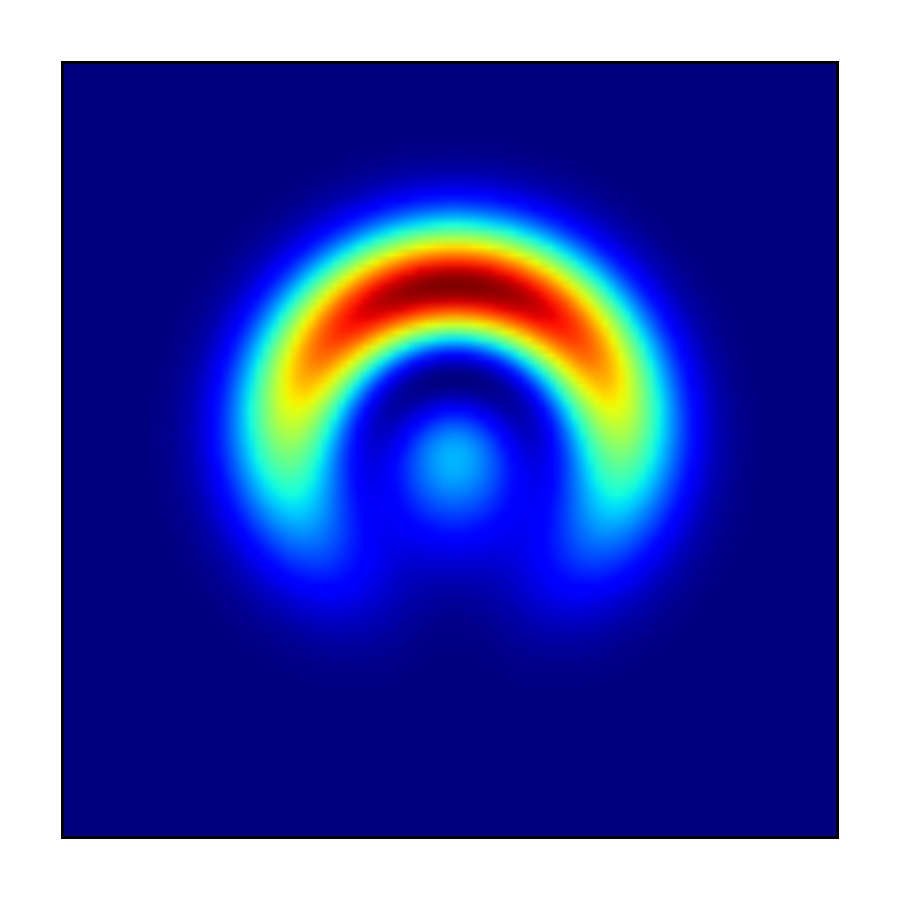} & \includepicture{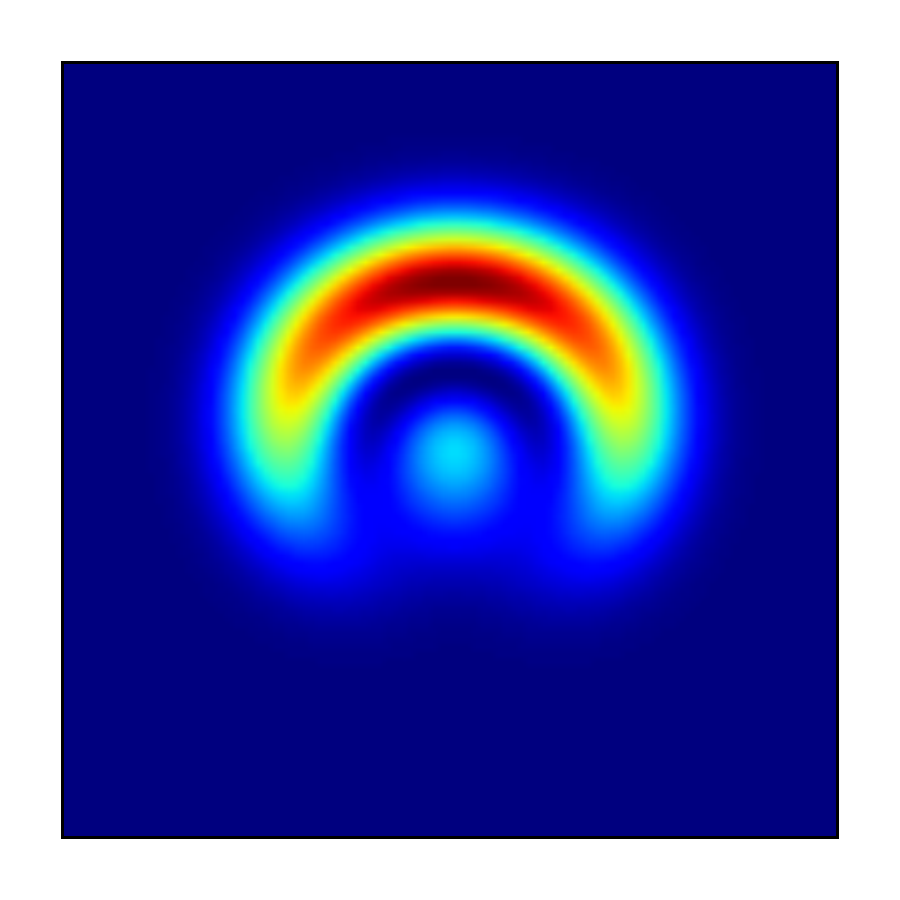} & \includepicture{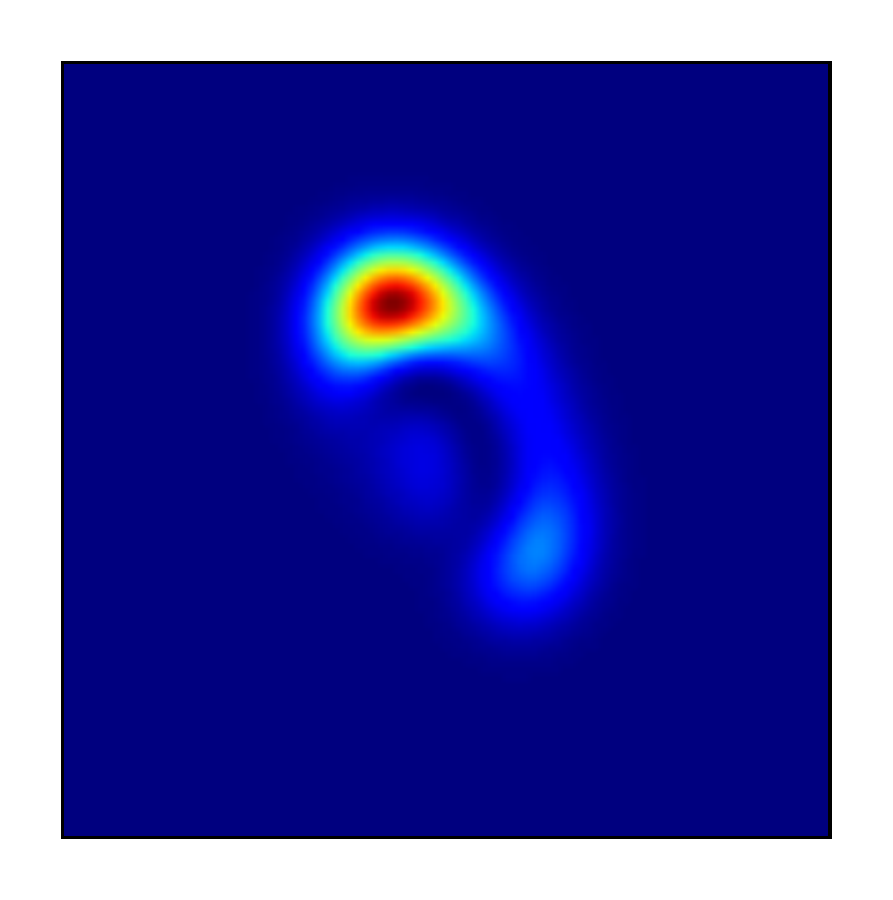} & \includepicture{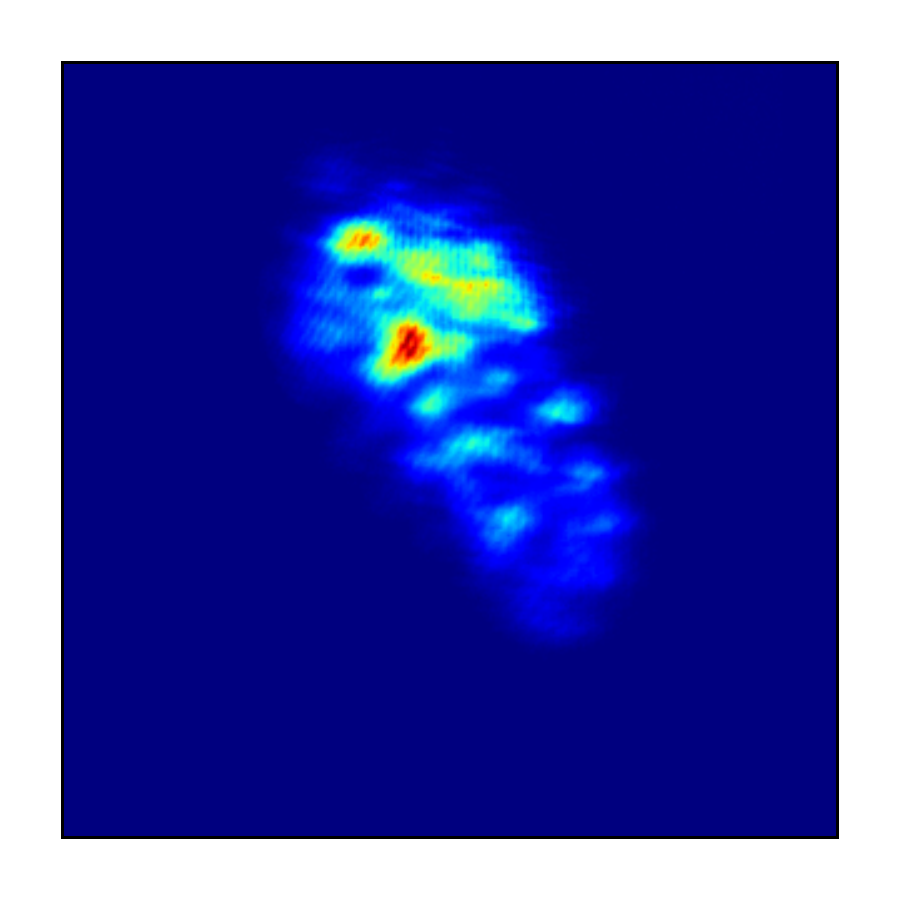}\\
$\frac{\ket{0}-\ket{1}}{\sqrt{2}}$ & \includepicture{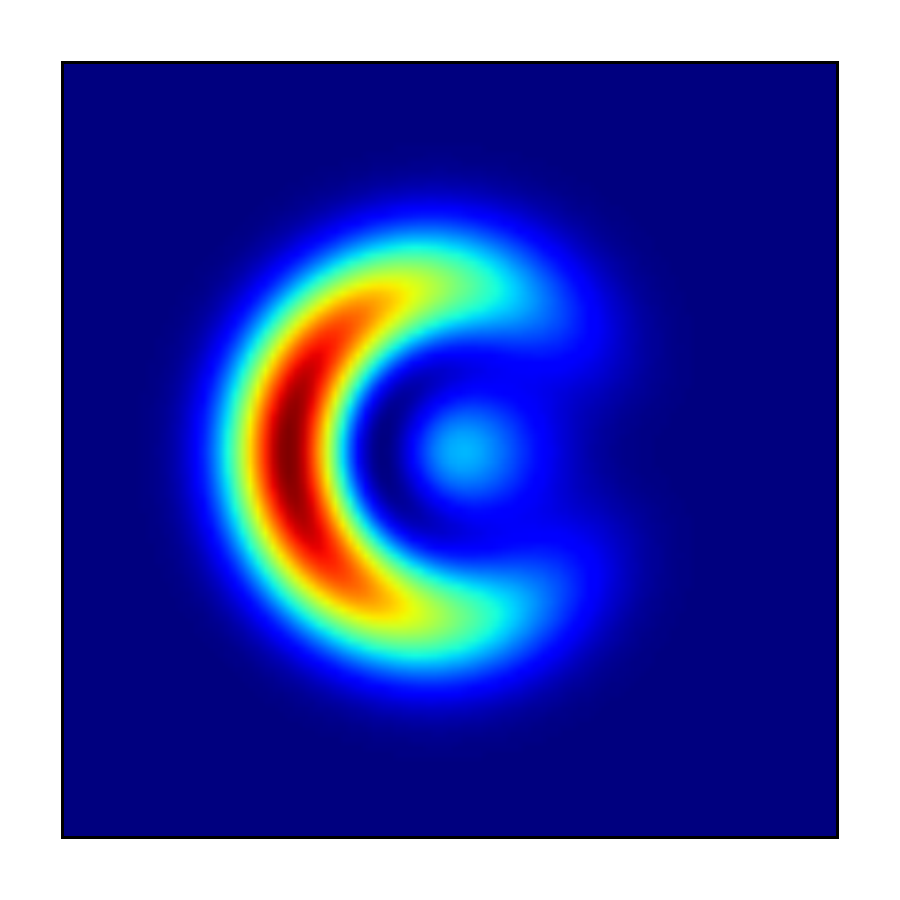} & \includepicture{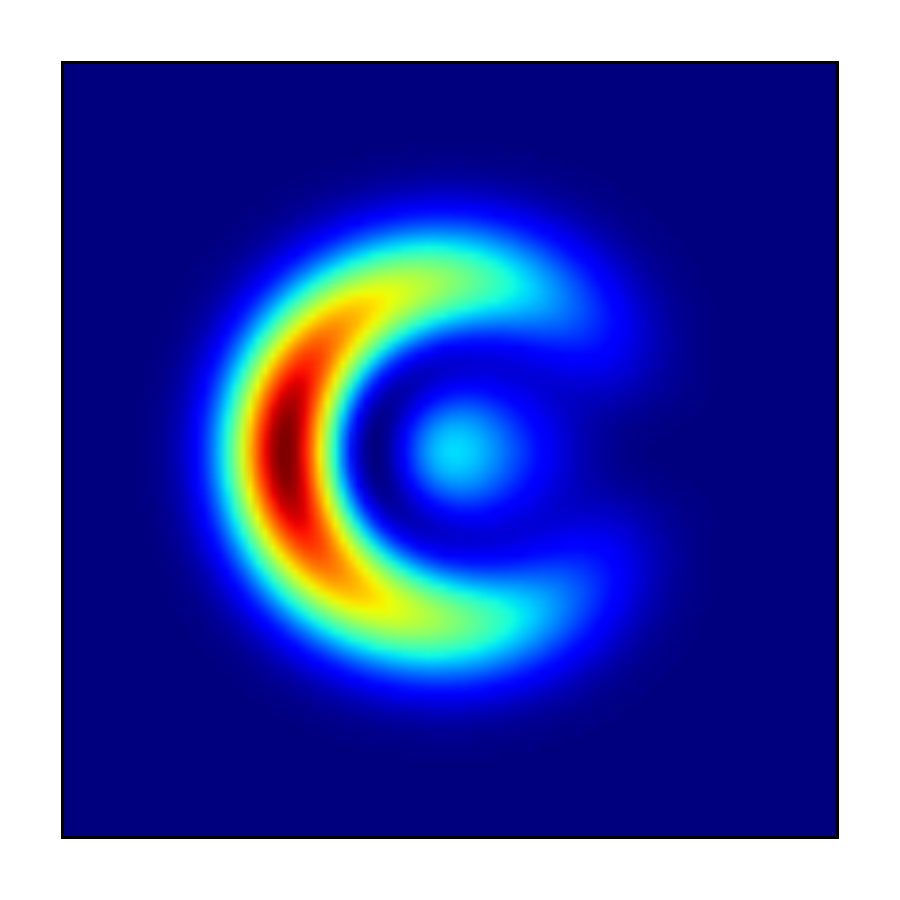} & \includepicture{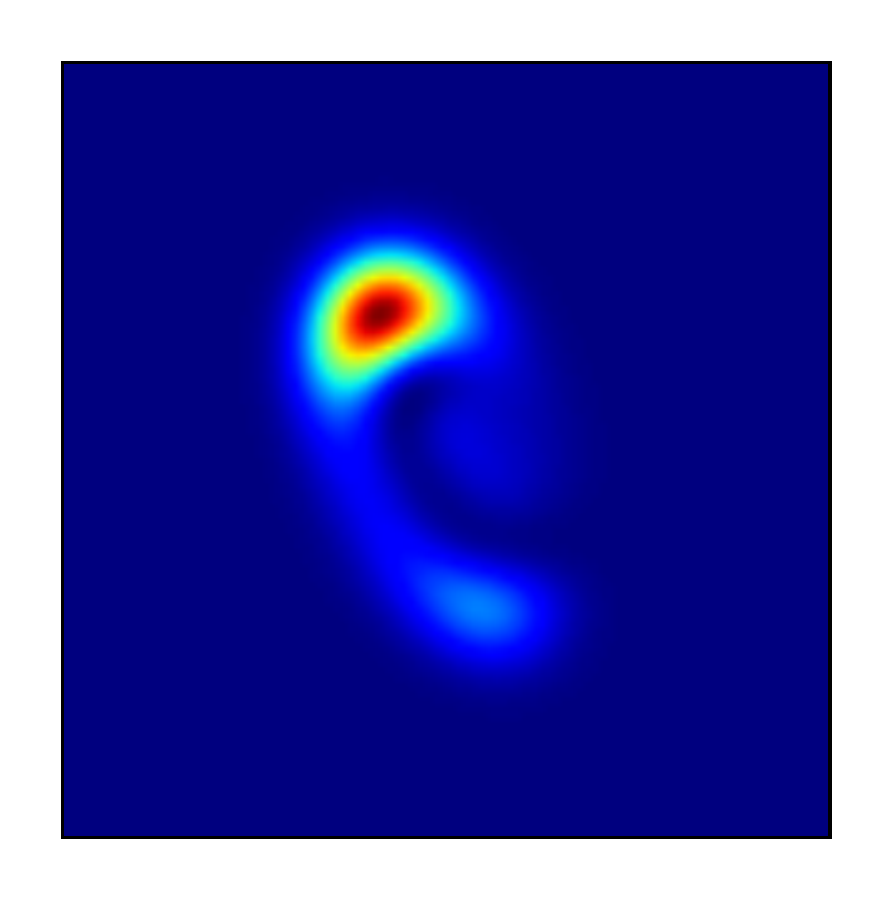} & \includepicture{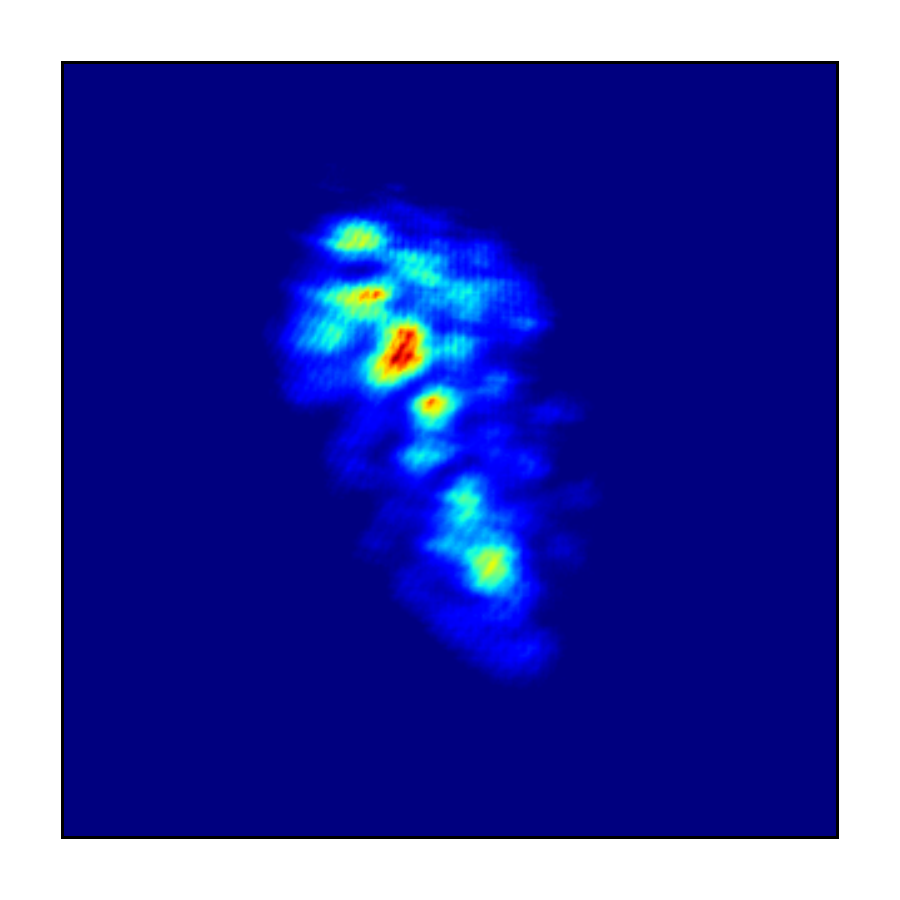}\\
\end{tabular} \\
\hline \\
\end{tabular}
\caption{Spacial probability distributions of the pointers for different initial states $\ket{0}$ and $\frac{\ket{0}-\ket{1}}{\sqrt{2}}$, weaknesses $\frac{\Delta}{\delta}$ and configurations.}
\label{fig:images}
\end{figure}

We use these BBDs in order to implement an approximation of the measurement described in \secref{sec:theory}. We do not take on the case of $n\rightarrow\infty$ here, but are content with the approximation provided by $n=6$. The results for the input states $\ket{0}$ and $\frac{1}{\sqrt{2}}\left(\ket{0}-\ket{1}\right)$ are shown in the second column of \figref{fig:images}. It can be seen that for very strong coupling, the images are quite different from the perfect case, but for weak couplings the approximation is quite good. We again compute the fidelity between the guessed state and the input state and plot it in \figref{fig:simFidelity}. It can be seen that the previous symmetry between different input states is now broken. This is expected since the symmetry is already broken in \eqnref{eq:Trotter} for finite $n$ since the operators on the right-hand-side, i.e. the couplings to $\sigma_z$ and $\sigma_x$, do not commute.

\begin{figure}
\subfloat[][]{\includegraphics[width=0.3\textwidth]{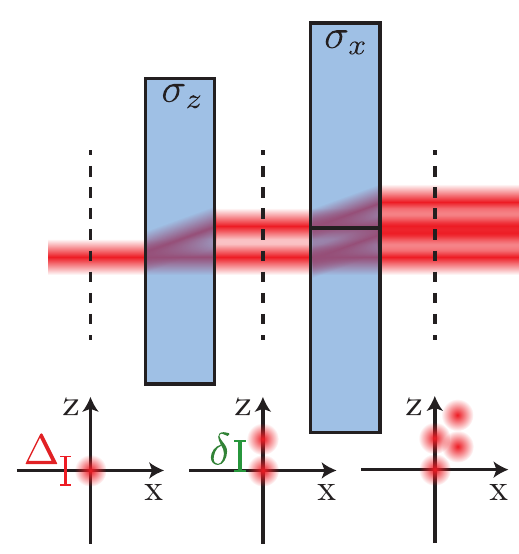}}\qquad \qquad
\subfloat[][]{\includegraphics[width=0.5\textwidth]{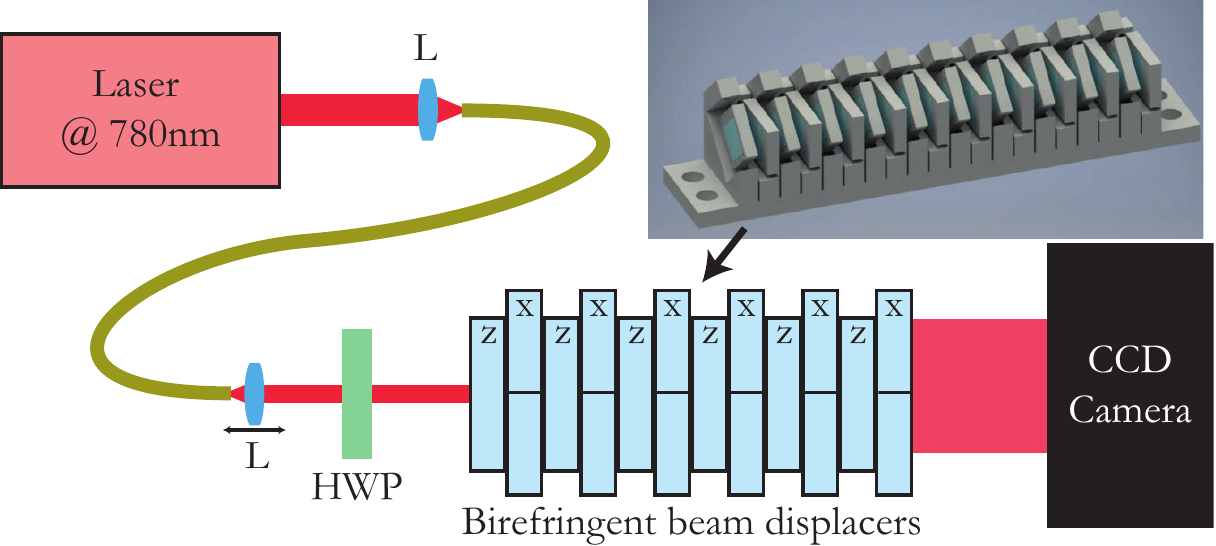}}
\caption{Experimental setup. a) Schematic depiction of the transformation performed by a birefringent crystal. b) Full experimental setup. L: lens; HWP: half-wave plate.}
\label{fig:experimentalSetup}
\end{figure}

As shown in \figref{fig:experimentalSetup}, the Trotter approximation can be implemented by alternating BBD crystals aligned at 0$^\circ$ and 45$^\circ$, to perform the measurement along $\sigma_z$ and $\sigma_x$, respectively.
In this configuration, the crystals do not implement exactly the transformations of Eq.(\ref{eq:Trotter}). Indeed the $Z$ crystal transformation is given by $e^{-i\frac{d}{2}\sigma_z\otimes\hat{p}_z\otimes\id}$, thus being exactly what we are looking for for $d=\frac{2\delta}{n}$, but the $X$ crystal transformation is described by $e^{i\frac{d}{\sqrt{2}}\sigma_x\otimes\left(\hat{p}_z\otimes\id+\id\otimes\hat{p}_x\right)}$; in other words it couples to both pointers\footnote{Alternatively, this can also be seen as each observable coupling to only one pointer with the pointers being non-orthogonal.}, as shown in \figref{fig:experimentalSetup}~(a). The differences of the probability distribution obtained with this approach and the one described previously, called Trotter 45$^\circ$ and Trotter 90$^\circ$, respectively, are shown in \figref{fig:images}. Note that transforming the images from the non-orthogonal reference frame into the orthogonal one does not yield the same probability distributions as before due to quantum interference effects. We can nonetheless perform this transformation and employ the same guessing strategy as before. The resulting fidelities are shown in \figref{fig:simFidelity}. While they are mostly lower than in the previous case, the second wind effect can still clearly be seen and the fidelities remain above $70\%$ in the interesting regime.\\

As presented in \figref{fig:experimentalSetup}~(b), to perform the experiment we use a continuous laser at 780\,nm coupled into a single-fiber to prepare the transverse mode as close as possible to a circular Gaussian state. At the output of the fiber, the beam waist is adjusted by a lens with a focal length of 6,24\,mm placed on a focus translation stage. The beam waist $\omega$ is directly related to the width $\Delta$ by the relation $\Delta  = \omega/2$. By changing the position of the lens, $\Delta$ can be adjusted from $0.05\,mm$ to $0.5\,mm$. The minimum value of $\Delta$ is limited by the diameter of the beam when it is focused on the camera placed at 30\,cm from the lens, whereas the maximum value is reached when which-path information is not negligible any more leading to a decrease of the interference. The initial state of the qubit is chosen using a half-wave plate, identifying horizontal polarization $\ket{H}$ and and vertical polarization $\ket{V}$ with $\ket{0}$ and $\ket{1}$, respectively. 

The BBDs are based on calcite crystal with a length of 5\,mm, which introduces a displacement $d$ of $500\,\mu m$ between the $H$ and $V$ polarization. At the output of the crystal the relative phase between the transmitted and displaced parts of the field depends on the length of the crystal and the incidence angle of the beam. Unfortunately, due to the imprecision on the cristal length, around 0.1\,mm, this phase is different for each crystal. As shown in the inset of \figref{fig:experimentalSetup}~(b), each crystal is placed on a tiltable support to set all these phases to zero.   

Since we only employ elements of linear optics, we do not need to work at the single-photon level; repeated single photon and a coherent laser lead to the same image. So, an 8 Mpixels CCD camera with a pixel size of 5.4/5.4$\,\mu m$ is used to directly measure the probability distribution at the output of the BBDs. The resulting images for different input polarizations and $\Delta/\delta$ are presented in the last column of \figref{fig:images}. We compute the resulting fidelities based on the same guesses as before and display them in \figref{fig:simFidelity}. The experimental fidelity follows the theoretical predictions. We see that increasing the weakness does not degrade the amount of extracted information.

%

\section{Conclusion and Outlook}

One can perform a simultaneous measurement of two non-commuting observables by coupling them to two independent pointers in the framework of weak von Neumann measurements. This type of measurement exhibits a yet to be fully understood second wind effect where weaker measurements can potentially reveal more information as measured by the average guessing fidelity. As a proof of principle, we implemented an approximation of this measurement on qubits encoded in the polarization of photons using pointers given by the transversal position of the same photons. The coupling was achieved using birefringent beam displacers. Due to experimental limitations, we were not able to conclusively demonstrate the second wind effect in our setup, but we observe that the amount of information is preserved when we increase the weakness.

The limiting factors in this first implementation were mostly technical. In order to get a better approximation of the intended measurement, it would be necessary to use a large amount of BBDs. Unfortunately this proves to be difficult. On the one hand, we are limited by the size of the BBDs and the camera aperture: if each BBD splits the beam by a distance $d$, then we need the size of the BBDs and the camera aperture to be larger than $nd$ plus the width of the beam in order for the relevant part of the beam to hit the camera and be measured. This can be improved by using narrower birefringent crystals, i.e. BBDs with a smaller displacement. On the other hand, the surface quality of the crystals introduced small random phases that cannot easily be compensated for, leading to errors that accumulate when transitioning towards using a larger amount of crystals. Furthermore, the displacements induced by the alternating BBDs were at a $45^\circ$ angle instead of the desired $90^\circ$. This problem could theoretically be overcome by introducing half wave plates after each crystal, allowing the displacements to be perpendicular.

The measurement setup we presented here basically performs tomography of the input state in the $x-z$ plane. This approach can be extended to other systems like electron beams with alternating Stern-Gerlach setups, coupling alternately to the vertical and horizontal spin directions. Moreover there is the possibility that a setup with two of these measurement devices reveals entanglement between two qubits. Some preliminary computations, which we do not present here, make us hopeful that this is indeed the case. Further investigations are required. It should also be pointed out that the measurements are weak, meaning that the measured state is not fully perturbed and could still contain some entanglement, thus being a case of one-shot non-entanglement-breaking entanglement detection. 

Finally, generalizing this type of measurement to higher dimensional systems remains an open problem.

\textit{Acknowledgments.---} The authors thank the atelier de mécanique of the University of Geneva for valuable technical assistance. This work was supported by the DIQIP project of the Swiss National Science Foundation (SNSF), the COST Action No.~MP1006 and the European Research Council (ERC MEC).

\bibliographystyle{apsrev4-1}
\bibliography{wmbib}

\newpage
\appendix

\section{Calculation for theoritcal measurement}
\label{app:meas}

Let $\ket{\psi_i}=\alpha\ket{0}+\beta\ket{1}$ be the initial state of the qubit. Let
\begin{align*}
\ket{\Phi_i}=\int_{-\infty}^{+\infty}\dd zG_{\Delta}(z)\ket{z}\otimes\int_{-\infty}^{+\infty}\dd x G_{\Delta}(x)\ket{x}
\end{align*}
with $G_{\Delta}(z)=\frac{1}{(2\pi\Delta^2)^{1/4}}e^{-\frac{z^2}{4\Delta^2}}$ be the initial state of the pointer. Using the relation between position and momentum, i.e. $\ket{z}=\frac{1}{\sqrt{2\pi}}\int_{-\infty}^{\infty}\dd p_ze^{-izp_z}\ket{p_z}$, we can equally write
\begin{align*}
\ket{\Phi_i}=\int_{-\infty}^{+\infty}\dd p_z\widetilde{G}_{\Delta}(p_z)\ket{p_z}\otimes\int_{-\infty}^{+\infty}\dd p_x \widetilde{G}_{\Delta}(p_x)\ket{p_x}
\end{align*}
with $\widetilde{G}_{\Delta}(p_z)=\frac{1}{\sqrt{2\pi}}\int_{-\infty}^{\infty}\dd ze^{-izp_z}G_{\Delta}(z)=\left(\frac{2\Delta^2}{\pi}\right)^{1/4}e^{-\Delta^2p_z^2}$.

We wish to calculate the state after the interaction, i.e.
\begin{align*}
\ket{\Psi_f}&=e^{-i\delta(\sigma_z\otimes\hat{p}_z+\sigma_x\otimes\hat{p}_x)}\ket{\psi_i}\ket{\Phi_i}\\
&=\int_{-\infty}^{\infty}\dd p_z\int_{-\infty}^{\infty}\dd p_x\widetilde{G}_{\Delta}(p_z)\widetilde{G}_{\Delta}(p_x)\left(e^{-i\delta(p_z\sigma_z+p_x\sigma_x)}\ket{\psi_i}\right)\ket{p_z}\ket{p_x}
\end{align*}
where we used that $\ket{p_z}$ and $\ket{p_x}$ are the Eigenfunctions of $\hat{p}_z$ and $\hat{p}_x$. By computing its Eigensystem we can simplify the operator
\begin{align*}
e^{-i\delta(p_z\sigma_z+p_x\sigma_x)}=\cos(\delta\sqrt{p_z^2+p_x^2})\id-i\frac{\sin(\delta\sqrt{p_z^2+p_x^2})}{\sqrt{p_z^2+p_x^2}}\left(p_z\sigma_z+p_x\sigma_x\right).
\end{align*}

We want to calculate the (subnormalized) postmeasurement state given that we got a specific measurement result $(z,x)$ for our pointer. Using $\braket{z}{p_z}=\frac{1}{\sqrt{2\pi}}e^{izp_z}$ we find
\begin{align*}
\ket{\psi_f(z,x)}&=\left(\id\otimes\bra{z,x}\right)\ket{\Psi_f}\\
&=\frac{1}{2\pi}\int_{-\infty}^{\infty}\dd p_z\int_{-\infty}^{\infty}\dd p_x\widetilde{G}_{\Delta}(p_z)\widetilde{G}_{\Delta}(p_x)e^{i(zp_z+xp_x)}\left(\cos(\delta\sqrt{p_z^2+p_x^2})\id-i\frac{\sin(\delta\sqrt{p_z^2+p_x^2})}{\sqrt{p_z^2+p_x^2}}\left(p_z\sigma_z+p_x\sigma_x\right)\right)\ket{\psi_i}.
\end{align*}
We expand $e^{i(zp_z+xp_x)}=\cos(zp_z+xp_x)+i\sin(zp_z+xp_x)$. Using symmetries and the linearity of integrals, we can see that several of the terms contribute 0. Note that $G_{\Delta}(p_z)$, $\cos(zp_z+xp_x)$ and $\cos(\delta\sqrt{p_z^2+p_x^2})$ are symmetric under the function $(p_z,p_x)\rightarrow(-p_z,-p_x)$ while $\frac{\sin(\delta\sqrt{p_z^2+p_x^2})}{\sqrt{p_z^2+p_x^2}}p_z$ and $\frac{\sin(\delta\sqrt{p_z^2+p_x^2})}{\sqrt{p_z^2+p_x^2}}p_x$ as well as $\sin(zp_z+xp_x)$ are antisymmetric. Denoting $\sqrt{p_z^2+p_x^2}$ by $p$ we are left with

\begin{align*}
&\ket{\psi_f(z,x)}=\frac{1}{2\pi}\int_{-\infty}^{\infty}\dd p_z\int_{-\infty}^{\infty}\dd p_x\widetilde{G}_{\Delta}(p_z)\widetilde{G}_{\Delta}(p_x)\left(\cos(zp_z+xp_x)\cos(\delta p)\id+\sin(zp_z+xp_x)\frac{\sin( \delta p)}{r}\left(p_z\sigma_z+p_x\sigma_x\right)\right)\ket{\psi_i}\\
&=\left(\frac{\Delta^2}{2\pi^3}\right)^{1/2}\int_{0}^{2\pi}\dd\phi\int_{0}^{\infty}\dd p pe^{-\Delta^2 p^2}\left(\cos\left(p(z\cos\phi+x\sin\phi)\right)\cos(\delta p)\id+\sin\left(p(z\cos\phi+x\sin\phi)\right)\sin(\delta p)\left(\cos\phi\sigma_z+\sin\phi\sigma_x\right)\right)\ket{\psi_i}
\end{align*}
where we fully switched to polar coordinates in the second line. Note that the $(z,x)$ dependence is exclusively of the form $a(\phi)=z\cos\phi+x\sin\phi$. The integrals over $p$ can be computed:
\begin{align*}
I_1(z,x)&=\left(\frac{\Delta^2}{2\pi^3}\right)^{1/2}\int_{0}^{2\pi}\dd\phi\int_{0}^{\infty}\dd p p^{-\Delta^2 p^2}\cos\left(pa(\phi)\right)\cos(\delta p) \\
&= \left(\frac{1}{2\pi\Delta^2}\right)^{1/2}-\frac{1}{2^{5/2}\pi^{3/2}\Delta^2}\int_0^{2\pi}\dd\phi\left((a(\phi)-\delta)\text{Dawson}\left(\frac{a(\phi)-\delta}{2\Delta}\right)+(a(\phi)+\delta)\text{Dawson}\left(\frac{a(\phi)+\delta}{2\Delta}\right)\right)\\
I_2(z,x)&=\left(\frac{\Delta^2}{2\pi^3}\right)^{1/2}\int_{0}^{2\pi}\dd\phi\int_{0}^{\infty}\dd p pe^{-\Delta^2 p^2}\sin\left(pa(\phi)\right)\sin(\delta p)\cos\phi \\
&= \frac{1}{2^{5/2}\pi^{3/2}\Delta^2}\int_0^{2\pi}\dd\phi\left(-(a(\phi)-\delta)\text{Dawson}\left(\frac{a(\phi)-\delta}{2\Delta}\right)+(a(\phi)+\delta)\text{Dawson}\left(\frac{a(\phi)+\delta}{2\Delta}\right)\right)\cos\phi\\
I_3(z,x)&=\left(\frac{\Delta^2}{2\pi^3}\right)^{1/2}\int_{0}^{2\pi}\dd\phi\int_{0}^{\infty}\dd p pe^{-\Delta^2 r^2}\sin\left(pa(\phi)\right)\sin(\delta p)\sin\phi \\
&= \frac{1}{2^{5/2}\pi^{3/2}\Delta^2}\int_0^{2\pi}\dd\phi\left(-(a(\phi)-\delta)\text{Dawson}\left(\frac{a(\phi)-\delta}{2\Delta}\right)+(a(\phi)+\delta)\text{Dawson}\left(\frac{a(\phi)+\delta}{2\Delta}\right)\right)\sin\phi
\end{align*}
Note that despite not being explicitly stated, $a(\phi)$ is also a function of $z$ and $x$. The Dawson function is given by $\text{Dawson}(x)=e^{-x^2}\int_0^x\dd ye^{y^2}$.

The subnormalized post-measurement state is thus
\begin{align*}
\ket{\psi_f(z,x)}=\left(I_1\id+I_2\sigma_z+I_3\sigma_x\right)\ket{\psi_i}
\end{align*}

For any fixed point $(z,x)$, these integrals can be computed numerically using standard solvers (we employed the quad solver of the scipy package of python), giving us the post-measurement state given that a specific measurement result $(z,x)$ was observed. Additionally, the probability of observing this measurement result is now simply given by the norm of the post-measurement state, i.e.
\begin{align*}
P(z,x|\psi_i)&=\braket{\psi_f(z,x)}{\psi_f(z,x)}\\
&=I_1^2+I_2^2+I_3^2+2I_1I_2\corr{\sigma_z}_{\ket{\psi_i}}+2I_1I_3\corr{\sigma_x}_{\ket{\psi_i}}
\end{align*}
with $\corr{\sigma_z}_{\ket{\psi_i}}=\bra{\psi_i}\sigma_z\ket{\psi_i}$. We used the calculations presented here in order to plot the figures given in the main text.

The guessing fidelity can now also be computed. Note that in polar coordinates, i.e. $z=r\cos\theta$, $x=r\sin\theta$, we have that $a(\phi)=r\cos\left(\theta-\phi\right)$. As mentioned in the main text, the average guessing fidelity is given by
\begin{align*}
F_{avg}(\ket{\psi_i})&=\int_{-\infty}^{+\infty}\dd z\int_{-\infty}^{+\infty}\dd xP(z,x|\psi_i)|\braket{\psi_{g}(z,x)}{\psi}|^2\\
&=\int_0^{2\pi}\dd\theta_g\left(\int_{0}^{+\infty}\dd rrP(z=r\cos\theta_g,x=r\sin\theta_g|\psi_i)\right)\cos^2\left(\frac{\theta_g-\theta_i}{2}\right).
\end{align*}
This can also be computed numerically.

\section{Calculations for Trotter measurements}
\label{app:trotter}
In this section we present the calculations behind our simulations and fidelities for the Trotter formula. While we could do the computation in a similar manner to the previous section, we can avoid the time intensive task of computing integrals by noting that the operators in the Trotter formula simply perform a translation of our Gaussian beams. Denoting by $\ket{\phi(z_0,x_0)}=\int\dd z\int\dd x G_{\Delta}(z-z_0)G_{\Delta}(x-x_0)\ket{z}\ket{x}$ a Gaussian beam centered at $(z_0,x_0)$, we have that
\begin{align*}
e^{-id\sigma_z\otimes\hat{p}_z}(\alpha\ket{H}+\beta\ket{V})\ket{\phi(z_0,x_0)}&=\alpha \ket{H}\ket{\phi(z_0-d,x_0)}+\beta\ket{V}\ket{\phi(z_0+d,x_0)}\\
e^{-id\sigma_x\otimes\hat{p}_x}(\alpha\ket{H}+\beta\ket{V})\ket{\phi(z_0,x_0)}&=\frac{\alpha+\beta}{2}(\ket{H}+\ket{V})\ket{\phi(z_0,x_0-d)}+\frac{\alpha-\beta}{2}(\ket{H}-\ket{V})\ket{\phi(z_0,x_0+d)}.
\end{align*}
We show this briefly for one case. We rewrite the $z$-part of the pointer state in the momentum basis via the Fourier transform
\begin{align*}
e^{-id\sigma_z\otimes\hat{p}_z}\ket{H}\ket{\phi(z_0,x_0)}&=\int\dd z\dd x\dd p_z G_{\Delta}(z)G_{\Delta}(x)e^{-izp_z}e^{-idp_z\sigma_z}\ket{H}\ket{p_z}\ket{x}\\
&=\int\dd z\dd x\dd p_z G_{\Delta}(z)G_{\Delta}(x)e^{-ip_z(z+d)}\ket{H}\ket{p_z}\ket{x}\\
&=\int\dd z\dd x\dd p_z G_{\Delta}(z-d)G_{\Delta}(x)e^{-ip_zz}\ket{H}\ket{p_z}\ket{x}\\
&=\ket{H}\ket{\phi(z_0-d,x_0)}.
\end{align*}
In the second line, we used that $\sigma_z\ket{H}=1\cdot\ket{H}$, in the third line we shifted the integration variable $z$ and for the last line we undid the Fourier transform to return to the position basis for $z$. The calculations for $\ket{V}$ is done similarly, using $\sigma_z\ket{V}=-1\cdot\ket{V}$. To do the calculation for $\sigma_x$, we simply switch to its Eigenbasis $\ket{\pm}=\frac{1}{\sqrt{2}}(\ket{0}\pm\ket{1})$ and do the same.\\

After $n$ applications of $e^{-id\sigma_x\otimes\hat{p}_x}e^{-id\sigma_z\otimes\hat{p}_z}$ we thus end up with an $n\times n$ grid of Gaussians of width $\Delta$, separated by a distance $d$ from its neighbors and with different amplitudes. The state at a particular location $(z,x)$ is then given by the overlap of these Gaussians with the corresponding amplitudes.\\

The simulation for our actual measurement at a 45$^\circ$ angle can be done the same way be realizing that
\begin{align*}
e^{-id\sigma_x\otimes(\cos\theta \hat{p}_z+\sin\theta\hat{p}_x)}(\alpha\ket{H}+\beta\ket{V})\ket{\phi(z_0,x_0)}&=\frac{\alpha+\beta}{2}(\ket{H}+\ket{V})\ket{\phi(z_0-d\cos\theta,x_0-d\sin\theta)}\\
&+\frac{\alpha-\beta}{2}(\ket{H}-\ket{V})\ket{\phi(z_0+d\cos\theta,x_0+d\sin\theta)}.
\end{align*}
A calculation analogous to above shows this. We still end up with an $n\times n$ grid of Gaussian beams (note that the grid no longer forms a square) whose overlap at a given point can easily be computed. The computation of the average guessing fidelity can then be done analogously to Appendix \ref{app:meas}.

\end{document}